\documentclass[prl,twocolumn,showpacs,amsmath,amssymb,superscriptaddress]{revtex4-2}

\usepackage{amsmath,amssymb,amsfonts,bm,color,graphicx,tabularx}

\usepackage[unicode=true,colorlinks=true]{hyperref}

\usepackage[etex=true,export]{adjustbox}
\usepackage{makecell}
\usepackage{multirow}
\usepackage{amsmath}
\usepackage{verbatim}

\hypersetup{linkcolor=blue, citecolor=blue,urlcolor=blue}
\usepackage{tabularx,multirow,array,diagbox}
\usepackage{adjustbox}

\newcommand{\beq}{\begin{equation}}
\newcommand{\eeq}{\end{equation}}
\newcommand{\beql}{\begin{equation*}}
\newcommand{\eeql}{\end{equation*}}
\newcommand{\beqn}{\begin{eqnarray}}
\newcommand{\eeqn}{\end{eqnarray}}

\begin{document}
\title{Hidden Zeeman Field in Odd-Parity Magnets: An Ideal Platform for Topological Superconductivity }

\author{Xun-Jiang Luo}
\email{xjluo@hmfl.ac.cn}
\affiliation{High Magnetic Field Laboratory, HFIPS, Chinese Academy of Sciences, Hefei, Anhui 230031, China}
\affiliation{Department of Physics, Hong Kong University of Science and Technology, Clear Water Bay, 999077 Hong Kong, China}
\author{Zi-Ting Sun}
\affiliation{Department of Physics, Hong Kong University of Science and Technology, Clear Water Bay, 999077 Hong Kong, China}
\author{Xilin Feng}
\affiliation{Department of Physics, Hong Kong University of Science and Technology, Clear Water Bay, 999077 Hong Kong, China}
\author{Mingliang Tian}
\email{tianml@hmfl.ac.cn}
\affiliation{High Magnetic Field Laboratory, HFIPS, Chinese Academy of Sciences, Hefei, Anhui 230031, China}
\author{K. T. Law}
\email{phlaw@ust.hk}
\affiliation{Department of Physics, Hong Kong University of Science and Technology, Clear Water Bay, 999077 Hong Kong, China}

\begin{abstract}
Odd-parity magnets (OPMs) have emerged as a fundamental class of unconventional magnetism, characterized by time-reversal-preserving non-relativistic spin splitting (NSS). Despite growing interest, the fundamental understanding of OPMs remains critically incomplete, as previous studies have focused exclusively on NSS while overlooking the intrinsically broken time-reversal symmetry ($\mathcal{T}$) inherent to magnetic order. In this Letter, we demonstrate that OPMs universally host a hidden Zeeman field rooted in this $\mathcal{T}$-breaking, which profoundly reshapes their band structure. Through an analytical study of an $f$-wave magnet model, we reveal that the NSS  microscopically originates from an emergent gauge field, manifesting as a real-space spin loop-current order in this model. Crucially, the large NSS (eV scale) enables conventional superconductivity to coexist robustly with the hidden Zeeman field, with the Zeeman splitting reaching hundreds of meV. This unique band structure establishes OPMs as an ideal platform for topological superconductors (TSCs), supporting large topological regions. Based on OPMs, we engineer a series of TSCs hosting distinct Majorana boundary modes, including unidirectional Majorana edge states.  Our work corrects a fundamental misconception about OPMs and establishes them as a versatile platform for field-free and robust TSCs.

\end{abstract}

\maketitle

\textit{Introduction.}---
Unconventional magnetism, characterized by symmetry-compensated magnetization and substantial non-relativistic spin splitting (NSS), has recently attracted intense interest \cite{YuanLinDing2020,Ma2021,Shao2021,ifmmode2022,LiuPengfei2022, Zhangsongbo2024,CheXiaobing2024, JiangYi2024, XiaoZhenyu2024,Maeda2024,
LiuYichen2024,Chen2025,Liu2025a,HuMengli2025,xilinfeng2025,Fukaya_2025,luoxun2026,luo2026unifiedsymmetry}. Among these systems, odd-parity magnets (OPMs) constitute a novel magnetic class which features Rashba-like spin splitting that is odd under momentum inversion \cite{HayamiSatoru2020,BirkHellenes2023,Brekke2024,zk69-k6b2,85fd-dmy8,YuPing2025,2025ziting,Sukhachov2025,Huang2025,2025li,2025zhu,zhuang2025,2025liu,2026arXiv260623806L}. Universal spin group symmetry criteria for OPMs have been established, leading to the identification of numerous candidate materials \cite{2025luospin,2025song,2026arXiv260307643L}. Experimentally, OPMs have been confirmed in both insulating and metallic compounds \cite{Song2025a,Yamada2025a}.
Despite these advances, the fundamental band structure of OPMs remains poorly understood. Previous studies have focused exclusively on the odd-parity NSS, which implicitly preserves time-reversal symmetry \(\mathcal{T}\) \cite{BirkHellenes2023,Yamada2025a}. However, \(\mathcal{T}\) is intrinsically  broken by the underlying magnetic order. Consequently,  an accurate description of the band structures of OPMs must simultaneously account for the \(\mathcal{T}\)-preserving NSS and the intrinsic \(\mathcal{T}\)-breaking. Addressing this issue is urgent for this rapidly advancing field and essential to unlock the full potential of OPMs for hosting novel phenomena.

The pursuit of topological superconductors (TSCs) is a central theme in condensed matter physics due to their promising applications in topological quantum computation \cite{Kitaev2001,Fu2008,Qi2008,Nayak2008,Hasan2010,Luo2021a,LuoMZM2024,xiaohong2024,zhang2024topological,Luobott,Luo2025vortex,2025Luoac}.
Conventionally, realizing a TSC requires the interplay of three ingredients: superconductivity, spin-orbit coupling (SOC), and $\mathcal{T}$ breaking, typically achieved by an external magnetic field \cite{Alicea_2012,Elliott2015,Pan2019}. This paradigm has been successfully implemented in various hybrid platforms, and great experimental progress has been achieved, notably in superconducting semiconductor systems \cite{Sau2010,Oreg2010,Lutchyn2010,Das2012,xinliu2016,Hell2017,Pientka2017,Lutchyn2018,Fornieri2019}. However, this mainstream strategy faces two major challenges. First, a threshold magnetic field is generally required to drive the topological phase, but strong fields tend to suppress superconductivity  \cite{ZhenZhu2021}. Second, the Zeeman splitting induced by the applied magnetic field is typically limited to the meV scale, resulting in a fragile topological phase that is highly sensitive to disorder \cite{chunxiaoliu2017}. These limitations motivate the search for intrinsic magnetic platforms that are compatible with superconductivity and can support large topological regions in parameter space, a goal that remains crucial and challenging.

\begin{figure}
\centering
\includegraphics[width=3.3in]{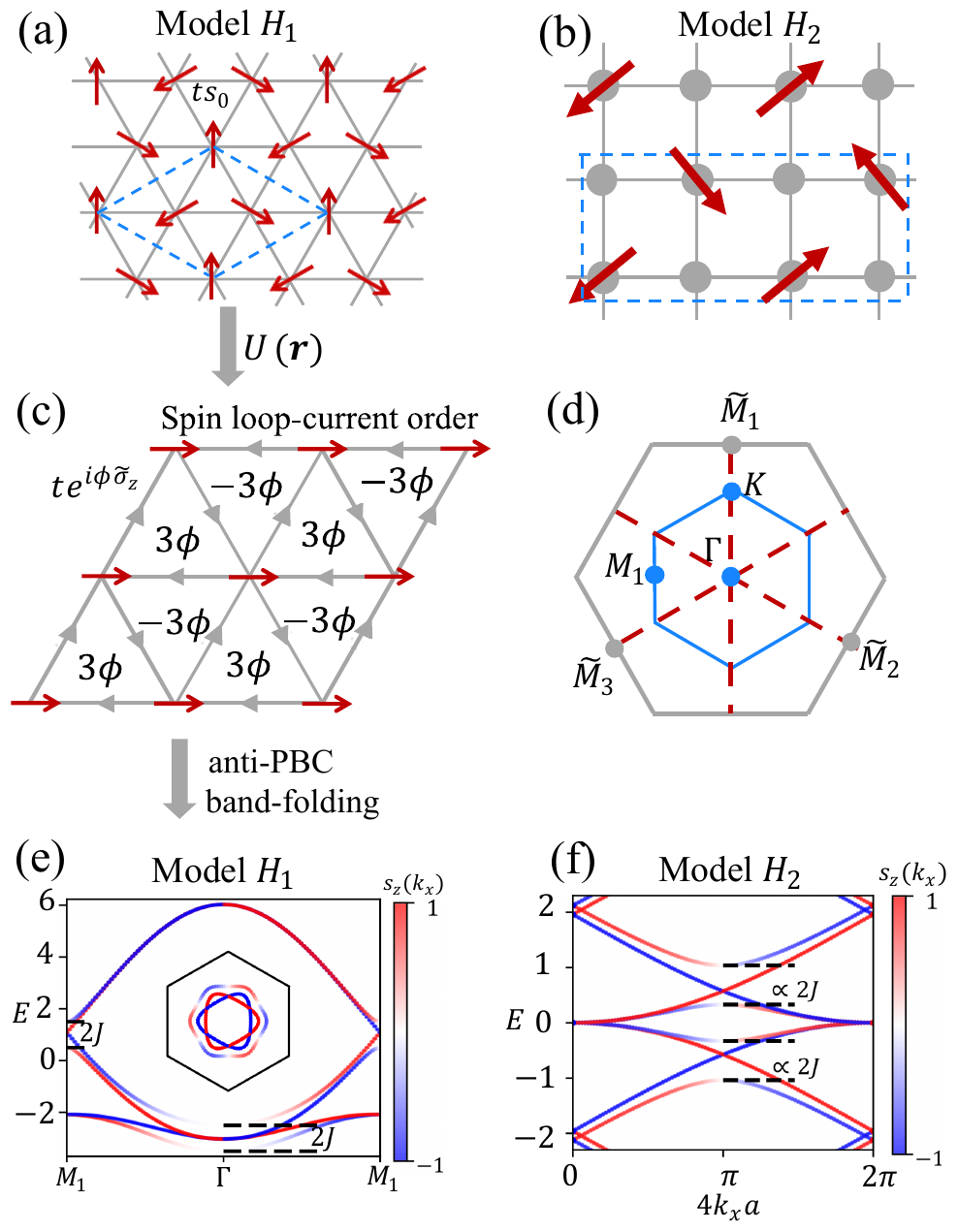}
\caption{ Band structures of OPMs. (a) and (b) Schematic  illustrations of models $H_1$ and $H_2$, respectively. The blue dashed lines denote the unit cells.
(c) Schematic  illustration of the mapping of $H_1$ to a spin loop-current order by $U(\bm r)$. (d) Brillouin zones of $H_1$ (blue) and the original triangular lattice (gray). (e) The bulk bands of model $H_1$; the inset plots its spin texture. (f) The bulk energy bands of model $H_2$.
In (c), $\phi=-\pi/3$, and anti-PBC stands for anti-periodic boundary conditions. In (e) and (f), $t=1$, $J=0.5$, and $\mu=-2.5$ for the inset plot in (e). }
\label{Fig1}
\end{figure}

In this Letter, we demonstrate that OPMs offer precisely such a platform. This capability stems from their unique band structure, which universally harbors both odd-parity NSS and a hidden Zeeman field (HZF). The HZF arises directly from intrinsic \(\mathcal{T}\)-breaking and manifests as the lifted Kramers degeneracy at time-reversal-invariant momenta [Figs.~\ref{Fig1}(e) and \ref{Fig1}(f)]. 
Through a representative $f$-wave magnet on a triangular lattice [{Fig.}~\ref{Fig1}(a)], we reveal the underlying band structure of OPMs by a gauge transformation. In the transformed frame, the magnetic order generates a gauge field, which
gives rise to a spin loop-current order in real space and the time-reversal-preserving NSS in momentum space. Meanwhile, an HZF naturally emerges, capturing the broken $\mathcal{T}$. Crucially, the large NSS (eV scale) pins the spin orientation, enabling conventional superconductivity to robustly coexist with the strong HZF (with Zeeman splitting reaching hundreds of meV). This unique band structure renders OPMs ideal platforms for TSCs, supporting large topological regions. Building on this foundation, we show that introducing $s$-wave superconducting pairing to OPMs can engineer various TSCs hosting distinct Majorana boundary modes, such as unidirectional Majorana edge states and spin-group-symmetry-protected Majorana Kramers pairs. Our work 
provides an accurate and complete description of the band structures of OPMs and establishes them as an ideal platform for robust and field-free TSCs.

\textit{HZF and emergent gauge field.}---
The NSS in OPMs can be effectively modeled by the term \(\bm{S}(\bm{k}) \cdot \bm{\sigma}\), where \(\bm{S}(\bm{k}) = -\bm{S}(-\bm{k})\) and \(\bm{\sigma} = (\sigma_x, \sigma_y, \sigma_z)\) denotes the Pauli matrices in spin space \cite{BirkHellenes2023}. However, such a term preserves the time-reversal symmetry \(\mathcal{T}\), whereas OPMs explicitly break \(\mathcal{T}\). A direct manifestation of this breaking is the lifting of Kramers degeneracy at time-reversal-invariant momenta, as shown in Figs.~\ref{Fig1}(e) and \ref{Fig1}(f). Although certain effective time-reversal operations such as \(\tilde{\mathcal{T}}_1 = \mathcal{T}\bm{\tau}\) or \(\tilde{\mathcal{T}}_2 = \mathcal{T}\sigma_z\) may be present in OPMs \cite{2025luospin}, they do not restore the full Kramers degeneracy. Specifically, \((\tilde{\mathcal{T}}_1)^2 = -e^{ik_x}\) (assuming that \(\bm{\tau}\) is a half-period translation along \(x\)) and \((\tilde{\mathcal{T}}_2)^2 = 1\); consequently, only \(\tilde{\mathcal{T}}_1\) can enforce a Kramers degeneracy, and only at  \(k_x = 0\) [Fig.~\ref{Fig1}(f)] \cite{Kudasov2024}. The generic absence of Kramers pairs at time-reversal-invariant momenta signals the existence of an HZF, as we will demonstrate.

To reveal the HZF in OPMs, we first consider a representative model \(H_1\), as schematically illustrated in Fig.~\ref{Fig1}(a). \(H_1\) describes an antiferromagnetic order defined on a triangular lattice with a $\sqrt{3}\times \sqrt{3}$ reconstruction and takes the form of
\begin{equation}
H_{1} = \sum_{\langle \bm{r}\bm{r}'\rangle,\sigma} t \, c^\dagger_{\bm{r}\sigma} c_{\bm{r}'\sigma}
+ J \sum_{\bm{r},\sigma\sigma'} \bm{m}(\bm{r}) \cdot c^\dagger_{\bm{r}\sigma} \bm{\sigma}_{\sigma\sigma'} c_{\bm{r}\sigma'},
\label{eq:realH}
\end{equation}
where \(c^\dagger_{\bm{r}\sigma}\) (\(c_{\bm{r}\sigma}\)) creates (annihilates) an electron with spin \(\sigma\) at the position \(\bm{r}\), \(\langle \bm{r}\bm{r}'\rangle\) runs over nearest-neighbor bonds with hopping amplitude \(t\), and the second term describes the exchange coupling to the local moment \(\bm{m}(\bm{r})\)  with strength \(J\). The magnetic texture is given by \(\bm{m}(\bm{r}) = (\cos(\bm{Q}\cdot\bm{r}+\phi), \sin(\bm{Q}\cdot\bm{r}+\phi), 0)\) with the antiferromagnetic propagation vector 
 \(\bm{Q} = \frac{4\pi}{3a}\bigl(\frac12,\frac{\sqrt3}{2}\bigr)\) and phase offset \(\phi = 7\pi/6\).
Enforced by spin-group symmetries, $H_1$ realizes an \(f\)-wave magnet in which the out-of-plane spin polarization takes the form $s_z(\bm k)\propto k_x(k_x^2 - 3k_y^2)$, while the in-plane components vanish, even though the real-space magnetic order is in-plane (see Appendix~\hyperlink{A}{A} for details).

The HZF in \(H_1\) becomes transparent after a unitary gauge transformation that rotates all moments to the \(x\)-direction: \(\tilde{H}_1 = U H_1 U^\dagger\), with \(U(\bm r) = e^{i (\bm{Q} \cdot \bm{r} + \phi) \sigma_z / 2}\). The transformed Hamiltonian reads
\beqn
\tilde{H}_1 &=& t \sum_{\bm{r},\sigma}\Bigl[ e^{-i\bm{Q}\cdot\bm{a}_1\tilde{\sigma}_z/2}\,\tilde{c}_{\sigma}^{\dagger}(\bm r)\tilde{c}_{\sigma}(\bm r+\bm{a}_1) \nonumber\\
&& + e^{-i\bm{Q}\cdot\bm{a}_2\tilde{\sigma}_z/2}\,\tilde{c}_{\sigma}^{\dagger}(\bm r)\tilde{c}_{\sigma}(\bm r+\bm{a}_2) \nonumber\\
&& + e^{-i\bm{Q}\cdot\bm{a}_3\tilde{\sigma}_z/2}\,\tilde{c}_{\sigma}^{\dagger}(\bm r)\tilde{c}_{\sigma}(\bm r+\bm{a}_3) + \text{h.c.} \Bigr] \nonumber\\
&& + J\sum_{\bm r,\sigma\sigma'} \tilde{c}_{\bm r\sigma}^\dagger (\tilde{\sigma}_x)_{\sigma\sigma'} \tilde{c}_{\bm r\sigma'},
\eeqn
where \(\bm{a}_1=(1,0)a\) (with \(a\) the lattice constant), \(\bm{a}_2=(-1/2,\sqrt{3}/2)a\), \(\bm{a}_3=-(\bm{a}_2+\bm{a}_1)\), $\tilde{\bm {\sigma}}=U\bm {\sigma }U^{\dagger}$, and \(\tilde{c}_{\bm r\sigma} =\sum_{\sigma'}U_{\sigma\sigma'}(\bm r) c_{\bm r\sigma'}\). In \(\tilde{H}_1\), the exchange coupling term becomes a uniform Zeeman field, which captures the broken $\mathcal{T}$. Moreover, an emergent gauge field appears: using \(\bm{Q}\cdot\bm{a}_1 = \bm{Q}\cdot\bm{a}_2 = \bm{Q}\cdot\bm{a}_3 = 2\pi/3\) (mod \(2\pi\)), each hopping term acquires a phase factor \(e^{-i\pi\tilde{\sigma}_z/3}\). The imaginary part of this hopping, preserving $\mathcal{T}$, gives rise to an odd-parity NSS.
Remarkably, these complex hoppings generate a loop flux on each triangular plaquette [Fig.~\ref{Fig1}(c)], which corresponds to a spin loop-current order in real space even though the charge current on each bond vanishes \cite{Shinjo2023}  (see Appendix~\hyperlink{B}{B} for details).
Since physical observables are basis-independent, this spin loop-current also exists in the original model \(H_1\).

\textit{Accurate band structures of OPMs.}---To accurately reveal the band structure of OPMs, we now further analyze the Bloch Hamiltonian of \(H_1\), which reads
\beqn
&&H_1=\begin{pmatrix}
-J(\sqrt{3}\sigma_x+\sigma_y)/2 & f(\bm k)\sigma_0 & f^{*}(\bm k) \sigma_0  \\
f^{*}(\bm k)\sigma_0 & J(\sqrt{3}\sigma_x-\sigma_y)/2 & f(\bm k)\sigma_0\\
f(\bm k) \sigma_0& f^{*}(\bm k)\sigma_0 & J\sigma_y\\
\end{pmatrix},\nonumber\\
\eeqn
where $f(\bm{k}) = t \sum_{i=1}^3 e^{i\bm{k}\cdot\bm{a}_i}$.  The energy bands of \(H_1(\bm k)\) are plotted in Fig.~\ref{Fig1}(e); the inset shows the value of $s_z(\bm k)$, which exhibits an \(f\)-wave pattern.

Remarkably,  $H_1(\bm k)$ can be block-diagonalized by performing two consecutive unitary transformations. The first transformation is
\beqn
U_1 = \operatorname{diag}\big( e^{i\theta_{\text{1}}\sigma_z}, -e^{i\theta_{\text{2}}\sigma_z}, -e^{i\theta_{\text{3}}\sigma_z} \big), 
\eeqn
with \(\theta_{\text{1}}=7\pi/12\), \(\theta_{\text{2}}=11\pi/12\), and \(\theta_{\text{3}}=\pi/4\). Applying \(H_1'(\bm{k}) = U_1 H_1(\bm{k}) U_1^\dagger\) yields
\beqn
H_1^{\prime}(\bm{k}) = 
\begin{pmatrix}
J\tilde{\sigma}_x & D(\bm{k}) & D^\dagger(\bm{k}) \\
D^\dagger(\bm{k}) & J\tilde{\sigma}_x & D(\bm{k}) \\
D(\bm{k}) & D^\dagger (\bm{k})& J\tilde{\sigma}_x
\end{pmatrix},
\eeqn
where \(D(\bm{k}) = -f(\bm{k}) e^{-i\frac{\pi}{3}\tilde{\sigma}_z}\). Notably, \(H_1^{\prime}\) restores translation symmetry due to the cyclic symmetry among sublattices, with the nearest-neighbor hopping becoming \(-te^{-i\pi/3\tilde{\sigma}_z}\). Thus, \(H_1^{\prime}(\bm k)\) corresponds to the system described by \(\tilde{H}_1(\bm r)\) with anti-periodic boundary conditions along the $\bm a_1$, $\bm a_2$, and $\bm a_3$ directions, which accounts for the global minus sign in \(D(\bm k)\) \cite{supp}.

The restored translation symmetry enables further block-diagonalization via a discrete Fourier transform:
\beqn
U_2 = \frac{1}{\sqrt{3}}
\begin{pmatrix}
\sigma_0 & \sigma_0 & \sigma_0 \\
\sigma_0 & \omega \sigma_0 & \omega^2 \sigma_0 \\
\sigma_0 & \omega^2 \sigma_0 & \omega \sigma_0
\end{pmatrix},
\eeqn
where \(\omega = e^{2\pi i/3}\). The transformation \(H_1^{\prime\prime}=U_2H_1^{\prime}U_2^{\dagger}\) gives a block-diagonal Hamiltonian \cite{supp}
\beqn
&&H_1^{\prime\prime}(\bm k) = \operatorname{diag}\big( h_0(\bm{k}), h_1(\bm{k}), h_2(\bm{k}) \big),\nonumber\\
&&h_p(\bm{k}) = J\tilde{\sigma}_x + \omega^{-p} D(\bm{k}) + \omega^{p} D^\dagger(\bm{k}), 
\eeqn
where \(p=0,1,2\) is the sublattice index. These blocks are connected by the magnetic propagation vector \(\bm{Q}\): using \(f(\bm{k}+\bm{Q}) = \omega f(\bm{k})\), we obtain
\beqn
h_2(\bm{k}) = h_0(\bm{k} + \bm{Q}), \quad h_1(\bm{k}) = h_0(\bm{k} - \bm{Q}).
\eeqn
Hence, \(H_1(\bm k)\) exactly maps to a two-band model \(h_0(\bm k)\); the energy bands of \(H_1(\bm k)\) are obtained by folding the bands of \(h_0(\bm k)\) into the magnetic Brillouin zone [Fig.~\ref{Fig1}(d)].
 \(h_0(\bm k)\) takes the form
\beqn
{h}_0(\bm k) = J\tilde{\sigma}_x - {f}_0(\bm k)\sigma_0 - {f}_z(\bm k)\tilde{\sigma}_z,
\eeqn
where \({f}_0(\bm k) = t\cos k_xa + 2t\cos (k_xa/2)\cos (\sqrt{3}k_ya/2)\) and \({f}_z(\bm k) = \sqrt{3}t\left(\sin k_xa - 2\sin (k_xa/2)\cos (\sqrt{3}k_ya/2)\right)\). The two-band model $h_0(\bm k)$ allows one to obtain the analytical expression for the spin expectation value $s_{x,y,z}(\bm k)$. In Appendix~\hyperlink{C}{C}, we show that $s_{x,y}(\bm k)=0$ and $s_z(\bm k)=f_z(\bm k)/\sqrt{J^2+f_z(\bm k)}$ (a certain band is considered).  Near the \(\Gamma\) point (\(\bm k=0\)), $s_z(\bm k)\propto k_x(k_x^2 - 3k_y^2)$, consistent with the symmetry analysis.

By the analytical study of \(H_1\), we uncover several key features that are universal to OPMs with coplanar magnetic order. First, the odd-parity NSS microscopically originates from an emergent gauge field induced by the magnetic order. In the rotated frame, the electron hopping generally becomes \(te^{i\theta\tilde{\sigma}_z/2}\), where \(\theta\) is the angle between neighboring coplanar magnetic moments. The imaginary part of these hoppings accounts for the time-reversal-preserving NSS.  Second, the energy scales of the NSS and HZF are set by the hopping amplitude \(t\) and the exchange coupling \(J\), respectively. Third, the HZF manifests identically in both the original lattice frame and the rotated frame: it lifts Kramers degeneracy at time-reversal-invariant momenta, accounting for the explicit \(\mathcal{T}\)-breaking. Specifically, in the rotated frame, the effective two-band model \(h_0(\bm k)\) exhibits a magnetic gap of \(2J\) opened by the HZF at time-reversal-invariant momenta. When folded back into the original magnetic Brillouin zone, these two bands gives rise to six bands. Consequently, only a subset of these bands possesses the HZF-induced gap at time-reversal-invariant momenta [Fig.~\ref{Fig1}(e)]; other apparent degeneracies at these momenta are mere artifacts of band folding (see details in the Supplemental Materials (SMs) \cite{supp}). Previous works focused on such nodal features but overlooked the band folding effect and the essential role of the HZF \cite{BirkHellenes2023,Yamada2025a}.

We note that the analytical study of \(H_1\) builds upon the helical magnetic order along the lattice bonds in real space. Such magnetic order is typically realized in helimagnets, which serve as minimal models for \(p\)-wave magnets \cite{Brekke2024}. These systems can be studied analogously: via a local unitary transformation, they can be mapped onto a two-band model in which both the NSS and HZF are explicitly revealed \cite{Martin2012,supp}. However, this mapping is not applicable to all OPMs. A notable exception is model \(H_2\), illustrated in Fig.~\ref{Fig1}(b), which realizes a \(p\)-wave magnet enforced by certain spin-group symmetries [see the SMs for details of the Hamiltonian and symmetry analysis]. Figure.~\ref{Fig1}(f) plots the energy bands of \(H_2\). Direct inspection reveals the lifting of Kramers degeneracy at \(k_x=\pi\) for certain bands, opening a magnetic energy gap proportional to \(2J\). However, the degeneracy at \(k_x=0\) remains, protected by the preserved symmetry \(\tilde{\mathcal{T}}_1\) in \(H_2\). The magnetic energy gap at \(k_x=\pi\) signals the existence of an HZF, which is further supported by the schematic decomposition shown in Fig.~\ref{Fig3}(c): \(H_2\) can be viewed as a stacking of one-dimensional (1D) helimagnets supplemented by local compensated Zeeman fields. This perspective reinforces that the coexistence of an odd-parity NSS and {the} HZF is a generic hallmark of OPMs.

\begin{figure}
\centering
\includegraphics[width=3.5in]{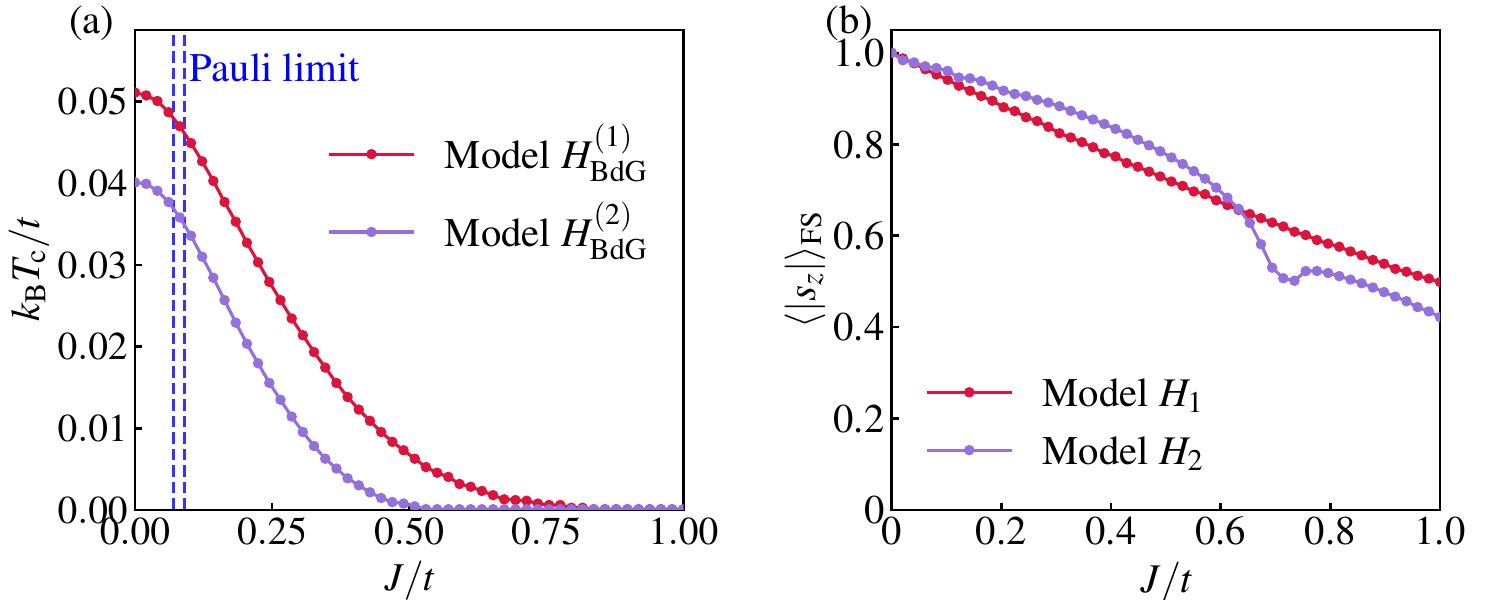}
\caption{(a) Self-consistently calculated critical temperature $T_c$ versus $J/t$ for models $H_{\text{BdG}}^{(1)}$ (red) and $H_{\text{BdG}}^{(2)}$ (purple). The dashed lines mark the respective Pauli limits. (b) Corresponding Fermi-surface-averaged out-of-plane spin polarization $\langle |s_z| \rangle_{\rm FS}$ as a function of $J/t$. Parameters: $t=1$, $U=1.5$, and $\mu=-1$.}
\label{FigTC}
\end{figure}

\textit{Compatibility between OPMs and superconductivity.}--- 
We now investigate the compatibility of OPMs with superconductivity by performing self-consistent mean-field calculations for models \(H_{1,2}\) with \(s\)-wave pairing. In these systems, the NSS pins the spin orientation along the \(z\)-direction, while the HZF lies in the \(xy\)-plane. This orthogonal configuration is reminiscent of Ising superconductors, where a large Ising SOC protects superconductivity against strong in-plane magnetic fields \cite{JMLu2015,Xi2016,Yingmingxie2020}. In OPMs, the energy scales of the NSS and HZF, determined by \(t\) and  \(J\),  can reach eV and hundreds of meV, respectively, in magnetic materials. Given these energy scales, we expect superconductivity to robustly coexist with OPMs.

To model the superconducting state, we introduce an attractive on-site interaction \(-U\) (with \(U>0\)) that mediates an\(s\)-wave pairing. The total Hamiltonian is given by \(H_{\text{SC}}^{(i)} = H_i - U \sum_{\bm r} c_{\bm r\uparrow}^{\dagger}c_{\bm r\downarrow}^{\dagger}c_{\bm r\downarrow}c_{\bm r\uparrow}\) for \(i=1,2\). Within mean-field theory, we define the order parameter \(\Delta = U\langle c_{\bm r\downarrow}c_{\bm r\uparrow}\rangle\), which leads to the Bogoliubov--de Gennes (BdG) Hamiltonian
\beqn
H_{\text{BdG}}^{(i)}(\bm k)=
\begin{pmatrix}
H_i(\bm k)-\mu & i\Delta\sigma_y\\
-i\Delta^{*}\sigma_y & -H_i^{*}(-\bm k)+\mu
\end{pmatrix},
\eeqn
where $\mu$ is the chemical potential.
The critical temperature \(T_c\) is obtained by numerically solving the linearized gap equation \(\partial^{2}\mathcal{F}_s(T,\Delta)/\partial\Delta^{2}|_{\Delta=0}=0\), where \(\mathcal{F}_s\) is the mean-field free energy density.

Figure.~\ref{FigTC}(a) shows the normalized critical temperature $T_c$ for models $H_{\text{BdG}}^{(1)}$ and $H_{\text{BdG}}^{(2)}$ as a function of $J/t$ at fixed $\mu=-1$. In both cases, superconductivity coexists with coplanar magnetic order over a wide range of $J$, far exceeding the Pauli limit marked by the dashed lines in Fig.~\ref{FigTC}(a). $T_c$ decreases continuously to zero, characteristic of a second-order transition.
To quantify this Ising-like protection, we introduce the Fermi-surface-averaged out-of-plane spin polarization $\langle |s_z| \rangle_{\rm FS}$ (defined in the SMs \cite{supp}). As shown in Fig.~\ref{FigTC}(b), $\langle |s_z| \rangle_{\rm FS}$ decreases monotonically with increasing $J$ before $T_c$ vanishes, consistent with the monotonically decreasing $T_c$ shown in Fig.~\ref{FigTC}(a). This correlation confirms that the NSS-induced protection is directly reflected by the magnitude of the spin polarization at the Fermi surface. Additional results for other chemical potentials are presented in the SMs~\cite{supp}.

\begin{figure}
\centering
\includegraphics[width=3.3in]{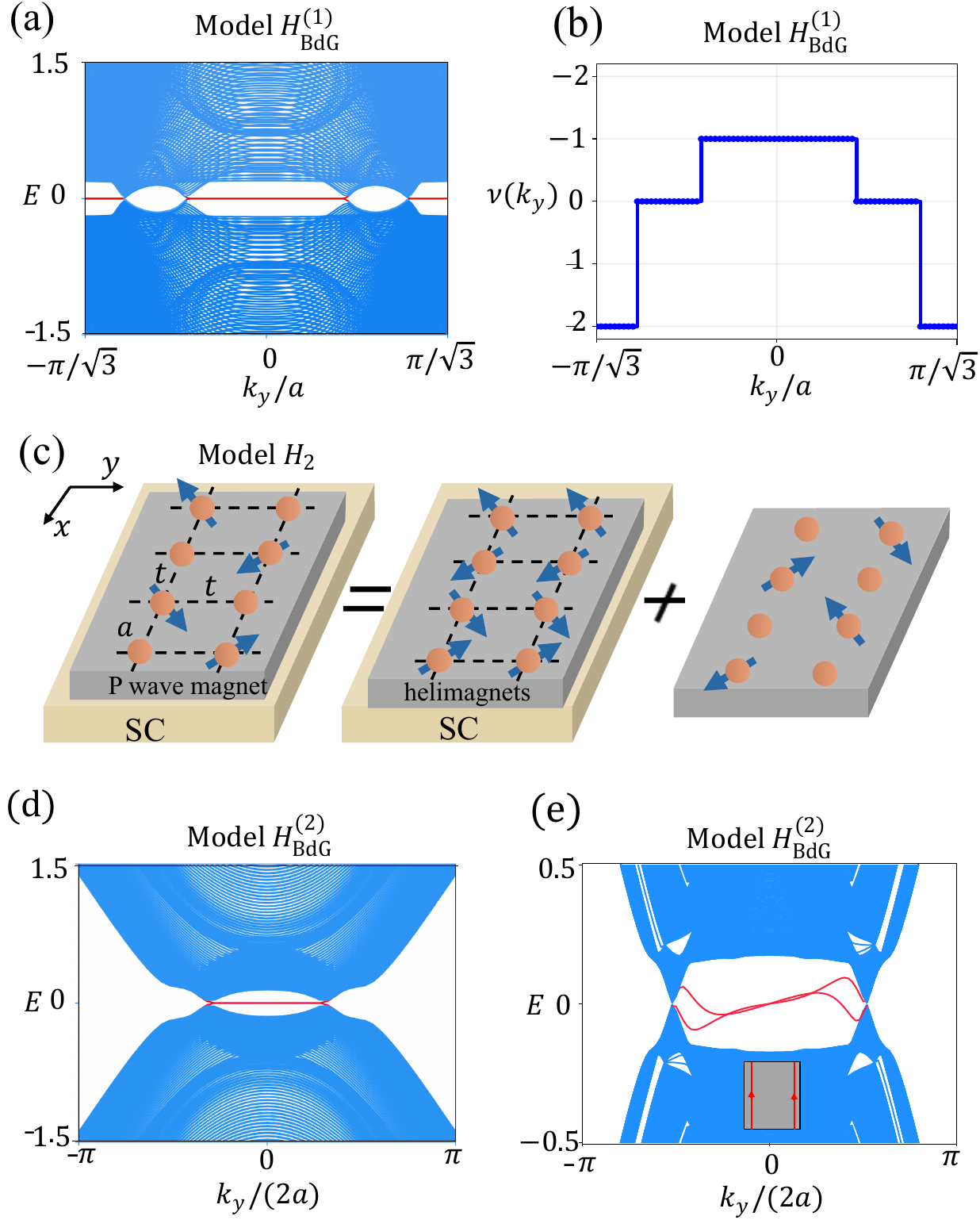}
\caption{(a) Energy spectrum of model $H_{\text{BdG}}^{(1)}$ in a nanowire geometry along the $y$ direction. (b) The momentum-dependent winding number of model $H_{\text{BdG}}^{(1)}$. (c) Schematic illustration of model $H_{\text{BdG}}^{(2)}$, mapping it to stacked superconducting 1D helimagnets with compensated Zeeman fields. (d) and (e) Energy bands of $H_{\text{BdG}}^{(2)}$ in a nanowire geometry along the $y$ direction without and with relativistic SOC, respectively. The model parameters are taken as $t=1$, $J=0.5$, $\Delta=0.2$, with $\mu=1$ for (a) and (b), and $\mu=3.5$ for (d) and (e).}
\label{Fig3}
\end{figure}

\textit{Engineering TSCs based on OPMs.}--- TSCs proximitized to noncollinear magnetic orders have been widely studied \cite{Choy2011,Nakosai2012,Vazifeh2013,Klinovaja2013,NadjPerge2013,Nadj-Perge2014,Soldini2023,Bazarnik2023}. Here, we focus on the OPM-based models $H_{\mathrm{BdG}}^{(1,2)}$, which respect the particle--hole symmetry $\mathcal{P} = \tau_x K$ and the effective time-reversal symmetry $\tilde{\mathcal{T}} = \mathcal{T}\tau_z\sigma_z$. These symmetries combine to yield the chiral symmetry $C = \mathcal{P}\tilde{\mathcal{T}} = \tau_y\sigma_x$, placing the systems in the BDI symmetry class. In two dimensions, $C$ can stabilize a nodal topology featuring Majorana flat bands, as shown in Figs.~\ref{Fig3}(a) and \ref{Fig3}(d) for $H_{\mathrm{BdG}}^{(1)}$ and $H_{\mathrm{BdG}}^{(2)}$, respectively. These gapless Majorana modes are characterized by the momentum-dependent winding number
\begin{equation}
\nu(k_y) = \frac{1}{2\pi}\int_{-\pi}^{\pi}dk_x\, \partial_{k_x}z(\bm{k}), \quad z(\bm{k}) = \frac{\det A(\bm{k})}{|\det A(\bm{k})|},
\end{equation}
where $A(\bm{k})$ is the off-diagonal block of the BdG Hamiltonian in the eigenbasis of $C$. Figure.~\ref{Fig3}(b) plots $\nu(k_y)$ for model $H_{\mathrm{BdG}}^{(1)}$, which precisely captures the topological transitions of the flat bands shown in Fig.~\ref{Fig3}(a).

The Majorana flat bands can be understood from the bulk band structure: at fixed $k_y=0$, the 1D slice exhibits a magnetic energy gap at time-reversal-invariant momenta. When the chemical potential lies within this gap and $s$-wave pairing is introduced, the corresponding 1D slice enters a TSC phase with a nonzero winding number. As $k_y$ varies, the effective parameters are renormalized and the topological condition is satisfied only within specific $k_y$ intervals, resulting in a topological nodal superconductor. For model $H_1$, since the normal-state Hamiltonian is exactly mapped to the two-band model $h_0(\bm{k})$, the BdG Hamiltonian reduces to an effective four-band model $H_{\mathrm{BdG}}^{(0)}$. Its energy gap closes when $f_z(\bm{k})=0$ (along the $\Gamma$--$\tilde{M}_{1,2,3}$ lines) and $J^2=\Delta^2+(f_0-\mu)^2$ \cite{supp}. These conditions define the topological region: $\mu \in \left[-t - \sqrt{J^2-\Delta^2},\; 3t + \sqrt{J^2-\Delta^2}\right]$, as detailed in the SMs \cite{supp}.

The Majorana flat bands in \(H_{\text{BdG}}^{(2)}\) admit another interpretation. As sketched in Fig.~\ref{Fig3}(c), the model $H_{\text{BdG}}^{(2)}$ can be decomposed into a stack of superconducting helimagnets with additional compensated Zeeman fields. A stack of superconducting helimagnets hosts nodal points and Majorana flat bands; the extra Zeeman fields merely shift the positions of the bulk nodes while preserving the flat bands \cite{supp}. Interestingly, when a relativistic SOC along the \(y\) direction is included, the Majorana flat bands become dispersive edge states, although the bulk nodal points remain \cite{supp}. The coexistence of gapless bulk states and dispersive edge states lifts constraints on the propagation direction of edge states, allowing scattering into the bulk. Consequently, edge states on opposite boundaries can propagate either in the same or opposite directions. When they flow in the same direction, unidirectional edge states emerge \cite{Wong2013}, as illustrated in the inset of Fig.~\ref{Fig3}(e). The SOC effect for model \(H_{\text{BdG}}^{(1)}\) is also studied in the SMs \cite{supp}.

\textit{Conclusion and discussion.}--- 
In summary, our work fundamentally reshapes the understanding of OPMs by revealing that their band structure universally hosts an HZF from intrinsic time-reversal-symmetry breaking, along with a time-reversal-preserving NSS from an emergent gauge field. This complete description addresses a critical gap in this rapidly advancing field. Harnessing this unique band structure, we demonstrate that OPMs provide an ideal platform for robust and field-free TSCs.

Although we have focused on models \(H_{1,2}\) as concrete examples to reveal the band structures of OPMs and demonstrate the engineering of robust TSCs, the underlying physics is general. This generality stems from the fundamental interplay between spin-group symmetry and time-reversal symmetry breaking in OPMs. In the SMs \cite{supp}, we further show that OPMs with coplanar or non-coplanar spin textures, distinct from the collinear cases of \(H_{1,2}\), can similarly realize robust TSCs, provided certain spin-group symmetry conditions are satisfied. Moreover, based on 1D helimagnets realizable in iron atom chains on a superconductor \cite{NadjPerge2013}, we demonstrate that both Majorana corner states and spin-group symmetry-protected Majorana Kramers pairs can be engineered, as detailed in the SMs \cite{supp}. These results establish OPMs as a versatile platform for exploring TSCs.

The use of OPMs as a platform for TSCs offers decisive advantages over conventional approaches relying on external magnetic fields. In standard semiconductor nanowire schemes, the topological phase requires \(m > \sqrt{\Delta^2 + \mu^2}\), where the Zeeman gap \(m\) is limited to the meV scale (e.g., \(\sim\)1.2 meV at 10 T for a typical \(g\)-factor of 2). The small Zeeman splitting makes the system extremely sensitive to nonuniform chemical potentials \cite{chunxiaoliu2017}. In stark contrast, OPMs intrinsically host both odd-parity NSS and an HZF. The latter can reach hundreds of meV in real materials, for instance, \(\sim\)130 meV in the \(p\)-wave magnet EuIn\(_2\)As\(_2\) \cite{supp,k9p4tfhd}, far exceeding the scale achievable with laboratory fields. This large intrinsic magnetic gap can coexist robustly with superconductivity due to the eV-scale NSS, dramatically expanding the topological regime in parameter space and significantly enhancing the robustness of Majorana boundary modes against disorder.

\section{Acknowledgments}
Xun-Jiang Luo thanks Ning Hao for helpful discussions. 
Mingliang Tian acknowledges the support of the National Key R\&D Program of the MOST of China (Grant No. 2022YFA1602603) and the Basic Research Program of the Chinese Academy of Sciences Based on Major Scientific Infrastructures (Grant
No. JZHKYPT- 2021-08). Xun-Jiang Luo  acknowledges the support of National Natural Science Foundation of China Grant No. 12604254.  K. T. L. acknowledges  the support of the Ministry of Science and Technology, China, The New Corner stone Foundation, and the Hong Kong Research Grants
Council through Grants No. MOST23SC01-A, No. RFS2021-6S03, No. C6053-23G, No. AoE/P-701/20, No. AoE/P-604/25R, No. 16309223, No. 16311424, and No. 16300325.

\bibliography{reference}

\clearpage

\appendix

\section{\large \textbf{End Matter}}

\renewcommand{\theequation}{A\arabic{equation}}
\setcounter{equation}{0}

\hypertarget{A}
{{\emph{Appendix A: Symmetry constraints on the spin textures}}}.---
In this appendix, we identify $H_1$ as an $f$-wave magnet by symmetry considerations.
$H_1$ respects the effective time-reversal symmetry $\tilde{\mathcal{T}}_2 = \mathcal{T}\sigma_z$, the spin-group symmetries $g_1 = \{U_z(2\pi/3)|\tau_1\}$ (where $\tau_1$ is a one-third-period translation along $x$), $g_2 = \{I||R_3\}$ (with $R_3$ denoting a spatial threefold rotation by $2\pi/3$ about the $z$ axis), and  $g_3=\{U_y(\pi)|M_x\}$, where $M_x$ is a mirror reflection $x\to -x$. These symmetries impose constraints on the spin expectation values $s_\alpha(\bm{k}) = \langle \psi_{\bm{k}} | \sigma_\alpha | \psi_{\bm{k}} \rangle$ ($\alpha = x, y, z$) as follows:
\begin{align}
\tilde{\mathcal{T}}_2&:\; s_z(\bm{k}) = -s_z(-\bm{k}), \quad s_{x,y}(\bm{k}) = s_{x,y}(-\bm{k}), \nonumber\\
g_1&:\; s_{x,y}(\bm{k}) = 0, \nonumber\\
g_2&:\; s_z(R_3\bm{k}) = s_z(\bm{k}), \nonumber\\
g_3&:\; s_z(M_x\bm{k}) = -s_z(\bm{k}),
\end{align}
where $|\psi_{\bm{k}}\rangle$ are the Bloch states at momentum $\bm{k}$.

To obtain the explicit form of $s_z(\bm{k})$, we expand it in angular harmonics. Writing $\bm{k} = (k\cos\theta, k\sin\theta)$ with $k=|\bm{k}|$, we have the general expansion
\beqn
s_z(\bm{k}) = \sum_{l=0}^{\infty} \left[ a_l(k) \cos(l\theta) + b_l(k) \sin(l\theta) \right],
\eeqn
where $a_l(k)$ and $b_l(k)$ are radial functions. The symmetry constraints restrict the allowed harmonics. The condition $s_z(\bm{k}) = -s_z(-\bm{k})$  requires $l$ to be odd. The condition $s_z(R_3\bm{k}) = s_z(\bm{k})$ demands invariance under $\theta \to \theta+2\pi/3$, which forces $l$ to be multiples of $3$. Together with odd $l$, the allowed harmonics are $l = 3, 9, 15, \dots$. Finally, the mirror symmetry $g_3$ imposes $s_z(M_x\bm{k}) = -s_z(\bm{k})$. Under $M_x$, $\theta \to \pi-\theta$. Using $\cos(3(\pi-\theta)) = -\cos(3\theta)$ and $\sin(3(\pi-\theta)) = \sin(3\theta)$, we find that $a_3(k)$ can be nonzero while $b_3(k)$ must vanish. Hence, the leading term at small $k$ is
\beqn
s_z(\bm{k}) = a_3(k) \cos(3\theta) + \mathcal{O}(k^9),
\eeqn
with $a_3(k) \propto k^3$. In Cartesian coordinates, $\cos(3\theta) = (k_x^3 - 3k_x k_y^2)/k^3$, yielding
\beqn
s_z(\bm{k}) \propto k_x(k_x^2-3k_y^2) + \mathcal{O}(k^9).
\eeqn
This $f$-wave angular dependence identifies $H_1$ as an $f$-wave magnet.

\renewcommand{\theequation}{B\arabic{equation}}
\setcounter{equation}{0}

\hypertarget{B}
{{\emph{Appendix B: Spin loop-current order }}}.---
In this appendix, we demonstrate that the transformed Hamiltonian \(\tilde{H}_1\) hosts a spin loop-current order, and that this order is physically equivalent to that in the original model \(H_1\). The model Hamiltonian \(\tilde{H}_1(\bm r)\) can be written as
\beqn
\tilde{H}_1 &=& \tilde{H}_{\text{hop}} + J\sum_{\bm r,\sigma\sigma'} \tilde{c}_{\bm r\sigma}^\dagger (\tilde{\sigma}_x)_{\sigma\sigma'} \tilde{c}_{\bm r\sigma'},\nonumber\\
\tilde{H}_{\text{hop}} &=& t \sum_{\bm{r},\sigma} \sum_{\bm{\delta}} \tilde{\omega}_{\bm{\delta}} \,\tilde{c}_{\bm{r}\sigma}^\dagger \tilde{c}_{\bm{r}+\bm{\delta},\sigma},
\eeqn
where the sum over \(\bm{\delta}\) runs over all six nearest-neighbor vectors of the triangular lattice: \(\bm{\delta} = \pm \bm{a}_1, \pm \bm{a}_2, \pm \bm{a}_3\). The phase factor is \(\tilde{\omega}_{\bm{\delta}} = e^{-i\sigma\pi/3}\) for \(\bm{\delta} = \bm{a}_i\) and \(\tilde{\omega}_{\bm{\delta}} = e^{i\sigma\pi/3}\) for \(\bm{\delta} = -\bm{a}_i\), ensuring \(\tilde{\omega}_{-\bm{\delta}} = \tilde{\omega}_{\bm{\delta}}^*\). Here \(\sigma = +1\) (\(\sigma = -1\)) denotes spin \(\uparrow\) (\(\downarrow\)).

\noindent
To derive the current operator, we consider the Heisenberg equation of motion (\(\hbar=1\))
\beqn
\frac{d}{dt} \tilde{n}_{\bm{r}} = i [\tilde{H}, \tilde{n}_{\bm{r}}] = i [\tilde{H}_{\text{hop}}, \tilde{n}_{\bm{r}}],
\eeqn
where the particle number operator is \(\tilde{n}_{\bm{r}} = \sum_{\tau} \tilde{c}_{\bm{r}\tau}^\dagger \tilde{c}_{\bm{r}\tau}\). The on-site Zeeman term commutes with \(\tilde{n}_{\bm{r}}\) because it is local and conserves the particle number.

Using the fermionic commutator identity with spin indices,
\beqn
[\tilde{c}_{i\sigma_1}^\dagger \tilde{c}_{j\sigma_2},\; \tilde{c}_{k\sigma_3}^\dagger \tilde{c}_{l\sigma_4}] = \delta_{jk}\delta_{\sigma_2\sigma_3} \tilde{c}_{i\sigma_1}^\dagger \tilde{c}_{l\sigma_4} - \delta_{il}\delta_{\sigma_1\sigma_4} \tilde{c}_{k\sigma_3}^\dagger \tilde{c}_{j\sigma_2},\nonumber\\
\eeqn
we evaluate the commutator for each bond. For a fixed site \(\bm{r}\) and a fixed nearest-neighbor vector \(\bm{\delta}\), the term \(\tilde{\omega}_{\bm{\delta}} \tilde{c}_{\bm{r}\sigma}^\dagger \tilde{c}_{\bm{r}+\bm{\delta},\sigma}\) gives a contribution \(-\tilde{\omega}_{\bm{\delta}} \tilde{c}_{\bm{r}\sigma}^\dagger \tilde{c}_{\bm{r}+\bm{\delta},\sigma}\) to \([\tilde{H}_{\text{hop}}, \tilde{n}_{\bm{r}}]\), while its Hermitian conjugate \(\tilde{\omega}_{\bm{\delta}}^* \tilde{c}_{\bm{r}+\bm{\delta},\sigma}^\dagger \tilde{c}_{\bm{r}\sigma}\) gives \(+\tilde{\omega}_{\bm{\delta}}^* \tilde{c}_{\bm{r}+\bm{\delta},\sigma}^\dagger \tilde{c}_{\bm{r}\sigma}\). Summing over all \(\bm{\delta}\) and spin \(\sigma\), we obtain
\beqn
[\tilde{H}_{\text{hop}}, \tilde{n}_{\bm{r}}] = t \sum_{\bm{\delta},\sigma} \left( -\tilde{\omega}_{\bm{\delta}} \tilde{c}_{\bm{r}\sigma}^\dagger \tilde{c}_{\bm{r}+\bm{\delta},\sigma} + \tilde{\omega}_{\bm{\delta}}^* \tilde{c}_{\bm{r}+\bm{\delta},\sigma}^\dagger \tilde{c}_{\bm{r}\sigma} \right).
\eeqn
Hence, the Heisenberg equation becomes
\beqn
\frac{d}{dt} \tilde{n}_{\bm{r}} = i t \sum_{\bm{\delta},\sigma} \left( -\tilde{\omega}_{\bm{\delta}} \tilde{c}_{\bm{r}\sigma}^\dagger \tilde{c}_{\bm{r}+\bm{\delta},\sigma} + \tilde{\omega}_{\bm{\delta}}^* \tilde{c}_{\bm{r}+\bm{\delta},\sigma}^\dagger \tilde{c}_{\bm{r}\sigma} \right).
\eeqn

To cast this as a continuity equation \(\frac{d}{dt} \tilde{n}_{\bm{r}} = -\sum_{\bm{\delta}} ( \tilde{J}_{\bm{r}\to\bm{r}+\bm{\delta}} - \tilde{J}_{\bm{r}+\bm{\delta}\to\bm{r}} )\), we define the charge current operator from \(\bm{r}\) to \(\bm{r}+\bm{\delta}\) as
\beqn
\tilde{J}_{\bm{r}\to\bm{r}+\bm{\delta}} = \frac{i t}{2} \sum_{\sigma} \left( \tilde{\omega}_{\bm{\delta}} \tilde{c}_{\bm{r}\sigma}^\dagger \tilde{c}_{\bm{r}+\bm{\delta},\sigma} - \tilde{\omega}_{\bm{\delta}}^* \tilde{c}_{\bm{r}+\bm{\delta},\sigma}^\dagger \tilde{c}_{\bm{r}\sigma} \right).
\eeqn
This operator is Hermitian and satisfies \(\tilde{J}_{\bm{r}+\bm{\delta}\to\bm{r}} = -\tilde{J}_{\bm{r}\to\bm{r}+\bm{\delta}}\). Similarly, the spin current operator for the \(z\)-component is defined by:
\beqn
\tilde{J}_{\bm{r}\to\bm{r}+\bm{\delta}}^{z} = \frac{i t}{2} \sum_{\sigma} \sigma \left( \tilde{\omega}_{\bm{\delta}} \tilde{c}_{\bm{r}\sigma}^\dagger \tilde{c}_{\bm{r}+\bm{\delta},\sigma} - \tilde{\omega}_{\bm{\delta}}^* \tilde{c}_{\bm{r}+\bm{\delta},\sigma}^\dagger \tilde{c}_{\bm{r}\sigma} \right).
\eeqn

The expectation value of the charge current in the ground state \(|\tilde{\Psi}_0\rangle\) is
\beqn
\langle \tilde{J}_{\bm{r}\to\bm{r}+\bm{\delta}} \rangle = \frac{i t}{2} \sum_{\sigma} \big( \tilde{\omega}_{\bm{\delta}} A_{\sigma}(\bm{\delta}) - \tilde{\omega}_{\bm{\delta}}^* A_{\sigma}^*(\bm{\delta}) \big),
\eeqn
where \(A_{\sigma}(\bm{\delta}) = \langle \tilde{c}_{\bm{r}\sigma}^\dagger \tilde{c}_{\bm{r}+\bm{\delta},\sigma} \rangle\) (translation invariance makes it independent of \(\bm{r}\)). The Hamiltonian \(\tilde{H}_1\) possesses the anti-unitary symmetry \(\Theta = i\tilde{\sigma}_x \mathcal{K}\) (with \(\mathcal{K}\) being complex conjugation), which satisfies \(\Theta \tilde{H}_1 \Theta^{-1} = \tilde{H}_1\) and exchanges spin up and down. This symmetry implies
\beqn
A_{-\sigma}(\bm{\delta}) = A_{\sigma}^*(\bm{\delta}).
\eeqn
Using this relation, one finds that the two spin contributions in \(\langle \tilde{J} \rangle\) cancel exactly, giving \(\langle \tilde{J}_{\bm{r}\to\bm{r}+\bm{\delta}} \rangle = 0\) for every bond.

For the spin current, the extra factor \(\sigma\) prevents cancellation. Using \(A_{-\sigma} = A_{\sigma}^*\) and  summing over \(\sigma\), we obtain
\beqn
\langle \tilde{J}_{\bm{r}\to\bm{r}+\bm{\delta}}^{z} \rangle = i t \big( e^{-i\pi/3} A_{+}(\bm{\delta}) - e^{i\pi/3} A_{+}^*(\bm{\delta}) \big),
\eeqn
which is generally nonzero and purely real. Consequently, each bond carries a persistent spin current.

We now consider an elementary triangular plaquette formed by sites \(\bm{r}_1,\bm{r}_2,\bm{r}_3\) in counter-clockwise order, with bonds along \(\bm{a}_1,\bm{a}_2,\bm{a}_3\) completing the triangle. The total spin current circulating around this plaquette is
\beqn
\mathcal{\tilde{J}}_{\triangle}^{z} = \tilde{J}_{\bm{r}_1\to\bm{r}_2}^{z} + \tilde{J}_{\bm{r}_2\to\bm{r}_3}^{z} + \tilde{J}_{\bm{r}_3\to\bm{r}_1}^{z}.
\eeqn
Because each bond carries the same magnitude of spin current, the sum is nonzero, indicating a spin loop-current order.

The original Hamiltonian \(H_1\) is related to \(\tilde{H}_1\) by the unitary transformation $U(\bm r)$. For any observable \(O\) in the original frame, its expectation value satisfies
\[
\langle \Psi_0 | O | \Psi_0 \rangle = \langle \tilde{\Psi}_0 | U O U^\dagger | \tilde{\Psi}_0 \rangle,
\]
where \(|\Psi_0\rangle = U^\dagger |\tilde{\Psi}_0\rangle\) is the ground state of \(H_1\). The physical spin current operator in the laboratory frame is \(J_{\bm{r}\to\bm{r}+\bm{a}_i}^{z,\text{lab}} = U^\dagger \tilde{J}_{\bm{r}\to\bm{r}+\bm{a}_i}^{z} U\). Since \(U\) is a spin rotation about the \(z\)-axis, it commutes with \(\sigma_z\) and therefore leaves the spin current operator unchanged. Consequently,
\beqn
\langle \Psi_0 | J_{\bm{r}\to\bm{r}+\bm{a}_i}^{z,\text{lab}} | \Psi_0 \rangle = \langle \tilde{\Psi}_0 | \tilde{J}_{\bm{r}\to\bm{r}+\bm{a}_i}^{z} | \tilde{\Psi}_0 \rangle.
\eeqn
Thus, the spin loop-current order exists in the original model as well, confirming its physical reality.

\renewcommand{\theequation}{C\arabic{equation}}
\setcounter{equation}{0}

\hypertarget{C}
{{\emph{Appendix C: Analytical calculations of the spin textures }}}.---
In this appendix, we analytically compute the spin expectation values $s_{x,y,z}(\bm k)$ for $H_1(\bm k)$. Recall that after the sequence of unitary transformations $U_1$ and $U_2$,  the Hamiltonian becomes block-diagonal:
\beqn
 H_1^{\prime\prime}(\bm k)&=&U_2U_1H_1(\bm k)U_1^\dagger U_2^\dagger\nonumber\\
&=&\operatorname{diag}\big(h_0(\bm k),h_1(\bm k),h_2(\bm k)\big),
\eeqn
where each $h_p(\bm k)$ ($p=0,1,2$) is a $2\times2$ matrix acting on the spin space. 
Concretely,
\beqn
h_p(\bm k) = J\tilde{\sigma}_x - f_0^{(p)}(\bm k)\sigma_0 - f_z^{(p)}(\bm k)\tilde{\sigma}_z,
\eeqn
with
\beqn
&&f_0^{(0)}(\bm k)=f_0(\bm k),\; f_z^{(0)}(\bm k)=f_z(\bm k),\nonumber\\
&&f_0^{(1)}(\bm k)=f_0(\bm k-\bm Q),\; f_z^{(1)}(\bm k)=f_z(\bm k-\bm Q),\nonumber\\
&&f_0^{(2)}(\bm k)=f_0(\bm k+\bm Q),\; f_z^{(2)}(\bm k)=f_z(\bm k+\bm Q),
\eeqn
It is noted that the tilde on $\tilde{\sigma}_x$ and $\tilde{\sigma}_z$ indicates that these Pauli matrices are expressed in the rotated spin basis.

Let $|\chi_p^\tau(\bm k)\rangle$ be an eigenstate of $h_p(\bm k)$ with band index $\tau=\pm$.  The eigenstate $|\psi_{\bm k}\rangle$ can be written as
\beqn
|\psi_{\bm k}\rangle = U_1^\dagger U_2^\dagger \; \big(|\chi_p^\tau\rangle \otimes |p\rangle\big),
\eeqn
where $|p\rangle$ denotes the sublattice index in the Fourier basis. 

We first consider the out-of-plane component $s_z(\bm k)$. Because $[U_{1,2},\sigma_z]=0$, we have
\beqn
s_z(\bm k) &=& \langle\psi_{\bm k}|\sigma_z|\psi_{\bm k}\rangle
=\langle\chi_{{p}}^\tau|\tilde{\sigma}_z|\chi_{{p}}^\tau\rangle \nonumber\\
&=& \tau\frac{f_z^{({p})}}{\sqrt{J^2+(f_z^{({p})})^2}}.
\eeqn
For the in-plane component $s_x(\bm k)$, we have
\beqn
s_x(\bm k) &=& \langle\psi_{\bm k}|\sigma_x|\psi_{\bm k}\rangle\nonumber\\
&=& \sum_{p^{\prime}} |(U_2^\dagger)_{p^{\prime}p}|^2 \langle\chi_{{p}}^\tau| (U_1)_{p^{\prime}}{\sigma}_x
(U_1^\dagger)_{p^{\prime}} |\chi_{{p}}^\tau\rangle\nonumber\\
& =& \frac{1}{3}\sum_{p^{\prime}} \langle\chi_{{p}}^\tau| \big[\cos(2\theta_{p^{\prime}}){\sigma}_x {-} \sin(2\theta_{p^{\prime}}){\sigma}_y\big] |\chi_{{p}}^\tau\rangle\nonumber\\
\eeqn
where {$(U_1)_1 = e^{i\theta_1\sigma_z}$, $(U_1)_2 = -e^{i\theta_2\sigma_z}$, and $(U_1)_3 = -e^{i\theta_3\sigma_z}$}, with  \(\theta_{\text{1}}=7\pi/12\), \(\theta_{\text{2}}=11\pi/12\), and \(\theta_{\text{3}}=\pi/4\).
We note that $\langle\chi_{{p}}^\tau|{\sigma}_y|\chi_{{p}}^\tau\rangle = 0$. Moreover, the trigonometric sums vanish:
\beqn
\sum_{p^{\prime}}\cos(2\theta_{p^{\prime}}) = \cos\frac{7\pi}{6}+\cos\frac{11\pi}{6}+\cos\frac{\pi}{2}
= 0.
\eeqn
Thus, we have $s_x(\bm k)=0$. Similarly, we have $s_y(\bm k)=0$.

\newpage

\begin{widetext}
\begin{center}
\begin{large}
\textbf{Supplemental Material for ``Hidden Zeeman Field in Odd-Parity Magnets: An Ideal Platform for Topological Superconductivity''}
\end{large}
\end{center}

This Supplemental Material includes the following seven sections:
(1) details of the band structures of model $H_1$;
(2) details of model $H_2$;
(3) topological superconductors (TSCs) based on models $H_{1,2}$;
(4) engineering TSCs based on 1D odd-parity magnets (OPMs);
(5) robustness of superconductivity;
(6) engineering Majorana corner states and Majorana Kramers
pairs;
(7) band structure of the $p$-wave magnet \texorpdfstring{EuIn\(_2\)As\(_2\)}{EuIn2As2}.

\subsection{A. Details of the band structures of model $H_1$}
The model $H_1$ describes a $120^\circ$ antiferromagnetic order
 on a triangular lattice with  a  $\sqrt{3}\times\sqrt{3}$ magnetic reconstruction. The magnetic unit cell contains three sublattices (A, B, and C), whose local magnetic moments are, respectively, given by
\[
\bm{m}_{\mathrm{A}} = \Bigl(-\frac{\sqrt3}{2},\, -\frac12,\, 0\Bigr),\qquad
\bm{m}_{\mathrm{B}} = \Bigl(\frac{\sqrt3}{2},\, -\frac12,\, 0\Bigr),\qquad
\bm{m}_{\mathrm{C}} = (0,\, 1,\, 0).
\]
The real‑space Hamiltonian reads
\begin{equation}
H_1(\bm r)=\sum_{\bm r,\sigma,\sigma'} J\,\bm m(\bm r)\cdot 
c_{\sigma}^{\dagger}(\bm r)\,\bm\sigma_{\sigma\sigma'}\,c_{\sigma'}(\bm r)
\;+\;t\sum_{\langle\bm r\bm r'\rangle,\sigma}
c_{\sigma}^{\dagger}(\bm r)c_{\sigma}(\bm r'),
\label{eq:H1_real}
\end{equation}
where \(c^\dagger_{\bm{r}\sigma}\) (\(c_{\bm{r}\sigma}\)) creates (annihilates) an electron with spin \(\sigma\) at the site \(\bm{r}\), \(\langle \bm{r}\bm{r}'\rangle\) runs over the nearest-neighbor pairs with hopping amplitude \(t\), and the first term describes the exchange coupling of strength \(J\) between the electron spin and the local magnetic moment \(\bm{m}(\bm{r})\). $\bm\sigma=(\sigma_x,\sigma_y,\sigma_z)$ are the Pauli matrices
and the spatially modulated magnetization is
$\bm m(\bm r)=\bigl(\cos(\bm Q\!\cdot\!\bm r+\phi),\,
\sin(\bm Q\!\cdot\!\bm r+\phi),0\bigr)$, 
where
$\bm Q=\frac{4\pi}{3a}\bigl(1/2,\sqrt3/2\bigr)$ and $\phi=7\pi/6$.

The band structures of  $H_1$ can be elucidated through a unitary transformation that aligns all local magnetic moments along a common direction. This unitary transformation can be chosen as  $U(\bm r) = e^{-i (\bm{Q} \cdot \bm{r} + \phi) \sigma_z / 2}$, which satisfies $U(\bm r) [\bm m(\bm r)\cdot \bm\sigma] U^\dagger(\bm r) = \tilde{\sigma}_x$. Under this transformation, the Hamiltonian becomes
\beqn
\tilde{H}_1(\bm r)&=&U(\bm r)H_1U^{\dagger}(\bm r)\nonumber\\
&=&\sum_{\langle \bm r\bm r'\rangle,\sigma\sigma'}\tilde{t}_{\sigma\sigma'} \, \tilde{c}_{\bm r\sigma}^\dagger \tilde{c}_{\bm r'\sigma'}  
+ J \sum_{\bm r,\sigma\sigma'} \tilde{c}_{\bm r\sigma}^\dagger (\tilde{\sigma}_x)_{\sigma\sigma'} \tilde{c}_{\bm r\sigma'},\nonumber\\
&=& t \sum_{\bm{r},\sigma}\Bigl[ e^{-i\bm{Q}\cdot\bm{a}_1\tilde{\sigma}_z/2}\,\tilde{c}_{\sigma}^{\dagger}(\bm r)\tilde{c}_{\sigma}(\bm r+\bm{a}_1)+ e^{-i\bm{Q}\cdot\bm{a}_2\tilde{\sigma}_z/2}\,\tilde{c}_{\sigma}^{\dagger}(\bm r)\tilde{c}_{\sigma}(\bm r+\bm{a}_2) \nonumber\\
&+& e^{-i\bm{Q}\cdot\bm{a}_3\tilde{\sigma}_z/2}\,\tilde{c}_{\sigma}^{\dagger}(\bm r)\tilde{c}_{\sigma}(\bm r+\bm{a}_3) + \text{h.c.} \Bigr]+J \sum_{\bm r,\sigma\sigma'} \tilde{c}_{\bm r\sigma}^\dagger (\tilde{\sigma}_x)_{\sigma\sigma'} \tilde{c}_{\bm r\sigma'},
\eeqn
where $\bm{a}_1=(1,0)a$, $\bm{a}_2=(-1/2,\sqrt{3}/2)a$, $\bm{a}_3=-(\bm{a}_2+\bm{a}_1)=(-1/2,-\sqrt{3}/2)$, 
$\tilde{\bm {\sigma}}=U\bm {\sigma }U^{\dagger}$, and \(\tilde{c}_{\bm r\sigma} =\sum_{\sigma'}U_{\sigma\sigma'}(\bm r) c_{\bm r\sigma'}\). Because $\bm{Q}\cdot\bm{a}_1 = \bm{Q}\cdot\bm{a}_2 = \bm{Q}\cdot\bm{a}_3 = 2\pi/3$ (mod $2\pi$), each hopping term carries the same phase $e^{-i\pi\tilde{\sigma}_z/3}$, which accounts the non-relativistic spin splitting (NSS) of $H_1$. Although this real-space picture intuitively shows the emergence of a Zeeman field and NSS, it does not fully account for the effects of $U(\bm r)$ on the boundary conditions. A more precise treatment is obtained by working directly with the Bloch Hamiltonian.

In the sublattice-spin space, the Bloch Hamiltonian $H_1(\bm k)$ can be written as
\beqn
H_1(\bm{k}) = 
\begin{pmatrix}
h_{\text{A}} / 2 & T_{\text{AB}}(\bm{k})\sigma_0 & T_{\text{AC}}(\bm{k})\sigma_0 \\
T_{\text{BA}}(\bm{k})\sigma_0 & h_{\text{B}} / 2 & T_{\text{BC}}(\bm{k})\sigma_0 \\
T_{\text{CA}}(\bm{k})\sigma_0 & T_{\text{CB}}(\bm{k})\sigma_0 & J\sigma_y
\end{pmatrix},
\eeqn
where $h_{\text{A}} = -J(\sqrt{3}\sigma_x + \sigma_y), \quad h_{\text{B}} = J(\sqrt{3}\sigma_x - \sigma_y), \quad T_{\text{AB}}(\bm{k}) = T_{\text{BC}}(\bm{k}) = T_{\text{AC}}^*(\bm{k}) = f(\bm{k})=t \sum_{i=1}^3 e^{i\bm{k}\cdot\bm{a}_i}$. The unitary transformation $U(\bm r)$ in momentum space corresponds to a block-diagonal matrix
\beqn
U_1 = \operatorname{diag}\big( U_{\text{A}}, U_{\text{B}}, U_{\text{C}} \big), \quad 
U_{\text{A}} = e^{i\frac{7\pi}{12}\sigma_z},\; U_{\text{B}} = e^{i\frac{11\pi}{12}\sigma_z},\; U_{\text{C}} = e^{i\frac{\pi}{4}\sigma_z}.
\eeqn
Applying $U_1$ yields the transformed Hamiltonian:
\beqn
H_1^{\prime}(\bm{k}) = U_1 H_1(\bm{k}) U_1^\dagger =
\begin{pmatrix}
J\tilde{\sigma}_x & f(\bm{k}) e^{-i\frac{\pi}{3}\tilde{\sigma}_z} & f^*(\bm{k}) e^{i\frac{\pi}{3}\tilde{\sigma}_z} \\[4pt]
f^*(\bm{k}) e^{i\frac{\pi}{3}\tilde{\sigma}_z} & J\tilde{\sigma}_x & f(\bm{k}) e^{i\frac{2\pi}{3}\tilde{\sigma}_z} \\[4pt]
f(\bm{k}) e^{-i\frac{\pi}{3}\tilde{\sigma}_z} & f(\bm{k})^* e^{-i\frac{2\pi}{3}\tilde{\sigma}_z} & J\tilde{\sigma}_x
\end{pmatrix}.
\eeqn
However, $H_1^{\prime}$ does not recover the cyclic symmetry in the sublattice indices, which corresponds to the translation symmetry of the original triangular lattice. To restore this symmetry, we apply a diagonal gauge transformation $U_2 = \operatorname{diag}\big( I, -I, -I \big)$,
which leads to
\beqn
H_1^{\prime\prime}(\bm{k}) = U_2 H_1'(\bm{k}) U_2^\dagger =
\begin{pmatrix}
J\tilde{\sigma}_x & -f(\bm{k}) e^{-i\frac{\pi}{3}\tilde{\sigma}_z} & -f^* (\bm{k})e^{i\frac{\pi}{3}\tilde{\sigma}_z} \\[4pt]
-f^*(\bm{k}) e^{i\frac{\pi}{3}\tilde{\sigma}_z} & J\tilde{\sigma}_x & f(\bm{k}) e^{i\frac{2\pi}{3}\tilde{\sigma}_z} \\[4pt]
f(\bm{k}) e^{-i\frac{\pi}{3}\tilde{\sigma}_z} & f^*(\bm{k}) e^{-i\frac{2\pi}{3}\tilde{\sigma}_z} & J\tilde{\sigma}_x
\end{pmatrix}.
\eeqn
Defining $D(\bm{k}) = -f(\bm{k}) e^{-i\frac{\pi}{3}\tilde{\sigma}_z}$ and noting that $e^{-i\pi\tilde{\sigma}_z}=-1$, we can write $H_1''(\bm{k})$ in a cyclically symmetric form:
\beqn
H_1^{\prime\prime}(\bm{k}) =
\begin{pmatrix}
J\tilde{\sigma}_x & D(\bm{k}) & D^\dagger(\bm{k}) \\
D^\dagger(\bm{k}) & J\tilde{\sigma}_x & D(\bm{k}) \\
D(\bm{k}) & D^\dagger(\bm{k}) & J\tilde{\sigma}_x
\end{pmatrix}.
\eeqn
Notably, $H_1^{\prime\prime}$ corresponds to the real-space Hamiltonian $\tilde{H}_1(\bm r)$ with anti-periodic boundary conditions along the $\bm a_1$, $\bm a_2$, and  $\bm a_3$ directions.

To block-diagonalize the Hamiltonian, we perform a discrete Fourier transform in sublattice space $\tilde{H}=U_3H_1^{\prime\prime}U_3^{\dagger}$, where
\[
U_3 = \frac{1}{\sqrt{3}}
\begin{pmatrix}
\sigma_0 & \sigma_0 & \sigma_0 \\
\sigma_0 & \omega \sigma_0 & \omega^2 \sigma_0 \\
\sigma_0 & \omega^2 \sigma_0 & \omega \sigma_0
\end{pmatrix}, 
\qquad \omega = e^{2\pi i/3}.
\]
It is noted that \(H_1^{\prime\prime}\) exhibits cyclic symmetry with matrix elements \((H_1^{\prime\prime})_{ij} = F_{(j-i)\bmod 3}\) where \(F_0 = J\tilde{\sigma}_x\), \(F_1 = D(\bm{k})\), \(F_2 = D^\dagger(\bm{k})\). The matrix elements of \(U_3\) can be written  as \((U_3)_{mi}= \omega^{mi}/\sqrt{3}\)  and those of its adjoint as \((U_3^\dagger)_{jn}= \omega^{-nj}/\sqrt{3}\)  (\(m,n,i,j=0,1,2\)). Therefore, 
the \((m,n)\) element of $\tilde{H}$ is
\beqn
\tilde{H}_{mn}= \sum_{i,j=0}^{2}(U_3)_{mi}(H_1^{\prime\prime})_{ij}(U_3^\dagger)_{jn}
               =\frac{1}{3}\sum_{i,j=0}^{2}\omega^{mi}(H_1^{\prime\prime})_{ij}\omega^{-nj}.
\eeqn
Using the cyclic form \((H_1^{\prime\prime})_{ij}=F_{(j-i)\bmod 3}\) and introducing the index shift \(k=j-i\) (so that \(j=i+k\)), the double sum decouples:
\beqn
\tilde{H}_{mn}= \frac{1}{3}\sum_{i=0}^{2}\omega^{mi}\Bigl[\sum_{k=0}^{2}F_{k}\,\omega^{-n(i+k)}\Bigr]
               =\frac{1}{3}\Bigl(\sum_{i=0}^{2}\omega^{(m-n)i}\Bigr)\Bigl(\sum_{k=0}^{2}\omega^{-nk}F_{k}\Bigr).
\eeqn
The summation over \(i\)  yields \(\sum_{i=0}^{2}\omega^{(m-n)i}=3\delta_{mn}\) and we have $\tilde{H}_{mn}= \delta_{mn}\sum_{k=0}^{2}\omega^{-nk}F_{k}$.  Consequently, \(\tilde{H}\) is diagonal in the sublattice indices, and 
the transformed Hamiltonian becomes:
\beqn
\tilde{H}(\bm{k}) = U_3 H_1^{\prime\prime}(\bm{k}) U_3^\dagger = \operatorname{diag}\!\big( h_0(\bm{k}), h_1(\bm{k}), h_2(\bm{k}) \big),
\eeqn
with each block given by
\beqn
h_m(\bm{k}) = J\tilde{\sigma}_x + \omega^{-m} D(\bm{k}) + \omega^{m} D^\dagger(\bm{k}), \qquad m = 0, 1, 2.
\eeqn

The three blocks are not independent. Using the property \(f(\bm{k}+\bm{Q}) = \omega f(\bm{k})\), one finds \(D(\bm{k}+\bm{Q})=\omega D(\bm{k})\) and \(D^\dagger(\bm{k}+\bm{Q})=\omega^{-1} D^\dagger(\bm{k})\). Consequently,
\beqn
h_0(\bm{k}+\bm{Q}) &=& J\tilde{\sigma}_x + \omega D(\bm{k}) + \omega^{-1} D^\dagger(\bm{k}) \nonumber\\
&=& J\tilde{\sigma}_x + \omega^{-2} D(\bm{k}) + \omega^{2} D^\dagger(\bm{k})\nonumber\\
&=& h_2(\bm{k}),
\eeqn
and similarly \(h_0(\bm{k}-\bm{Q}) = h_1(\bm{k})\). Thus, the full spectrum is captured by a single effective two‑band Hamiltonian, which we take as
\beqn
h_0(\bm{k}) &=& J\tilde{\sigma}_x - f(\bm{k}) e^{-i\frac{\pi}{3}\tilde{\sigma}_z} - f^{*}(\bm{k}) e^{i\frac{\pi}{3}\tilde{\sigma}_z}\nonumber\\
&=&J\tilde{\sigma}_x-f_0(\bm k)\sigma_0-f_z(\bm k)\tilde{\sigma}_z,
\label{s15}
\eeqn
where \({f}_0(\bm k) = t\cos k_xa + 2t\cos (k_xa/2)\cos (\sqrt{3}k_ya/2)\) and \({f}_z(\bm k) = \sqrt{3}t\left(\sin k_xa - 2\sin (k_xa/2)\cos (\sqrt{3}k_ya/2)\right)$.

To visualize the band folding explicitly, we plot in Fig.~\ref{fig:folding} the energy bands of the two-band model $h_0(\bm{k})$ in the extended momentum range [Fig.~\ref{fig:folding}(a)] and the resulting folded bands in the magnetic Brillouin zone [Fig.~\ref{fig:folding}(b)]. The three sectors (labeled Sector 1, 2, and 3) correspond to the three magnetic zones centered at $0$, $\pm\bm{Q}$, respectively. Upon folding, they combine to produce the six bands of $H_1(\bm{k})$. It is evident that only a subset of bands carries the hidden Zeeman splitting at the $\Gamma$ point; the other degeneracies are artifacts of band folding.

\begin{figure*}
\centering
\includegraphics[width=0.85\textwidth]{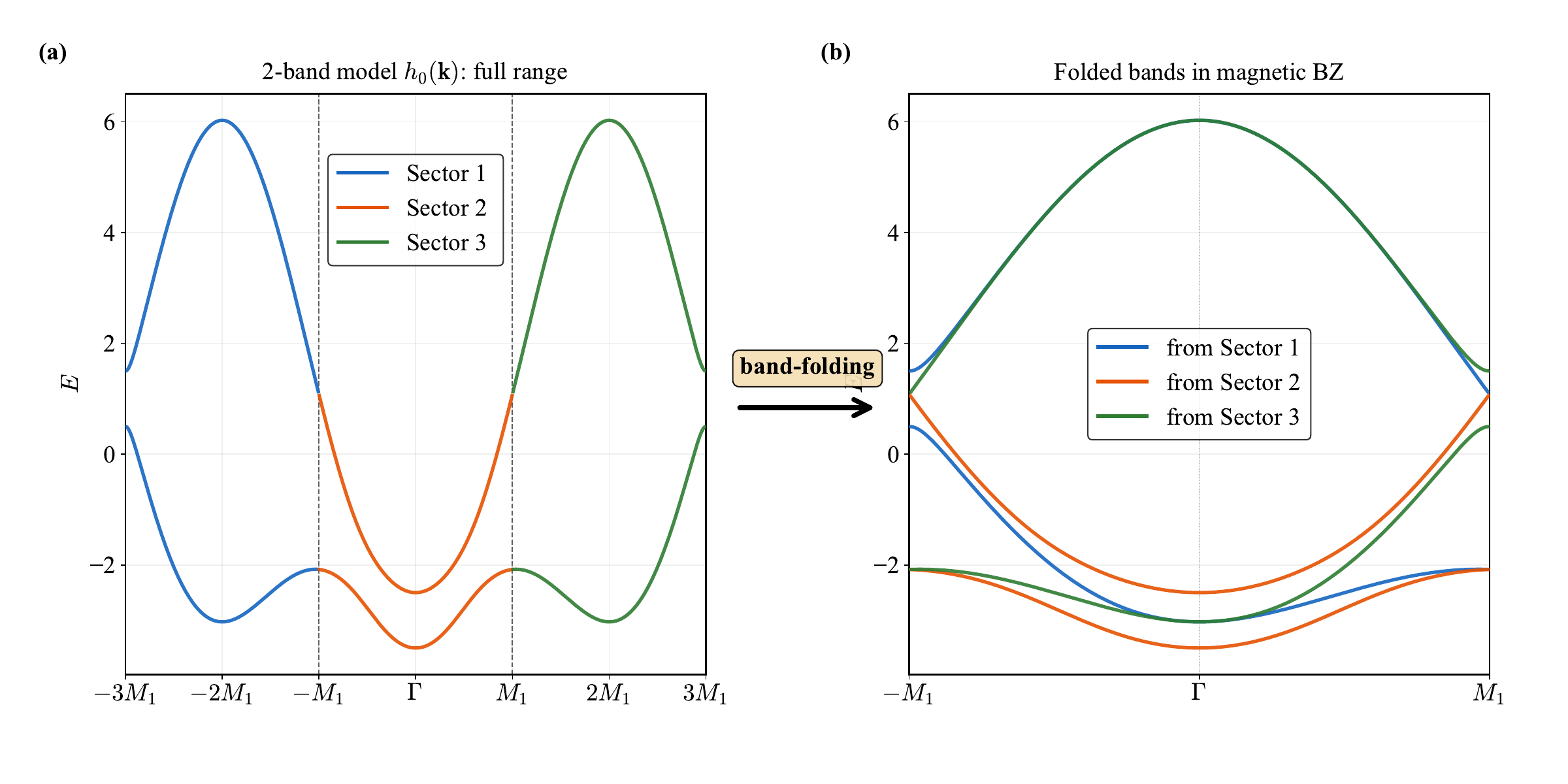}
\caption{(a) Energy bands of the two-band model $h_0(\bm{k})$ in the extended momentum range, showing three sectors (blue, orange, and green curves). (b) The folded bands in the magnetic Brillouin zone obtained from the three sectors via band folding. The parameters are $t=1$ and $J=0.5$.}
\label{fig:folding}
\end{figure*}

{
\begin{figure*}
\centering
\includegraphics[width=6.in]{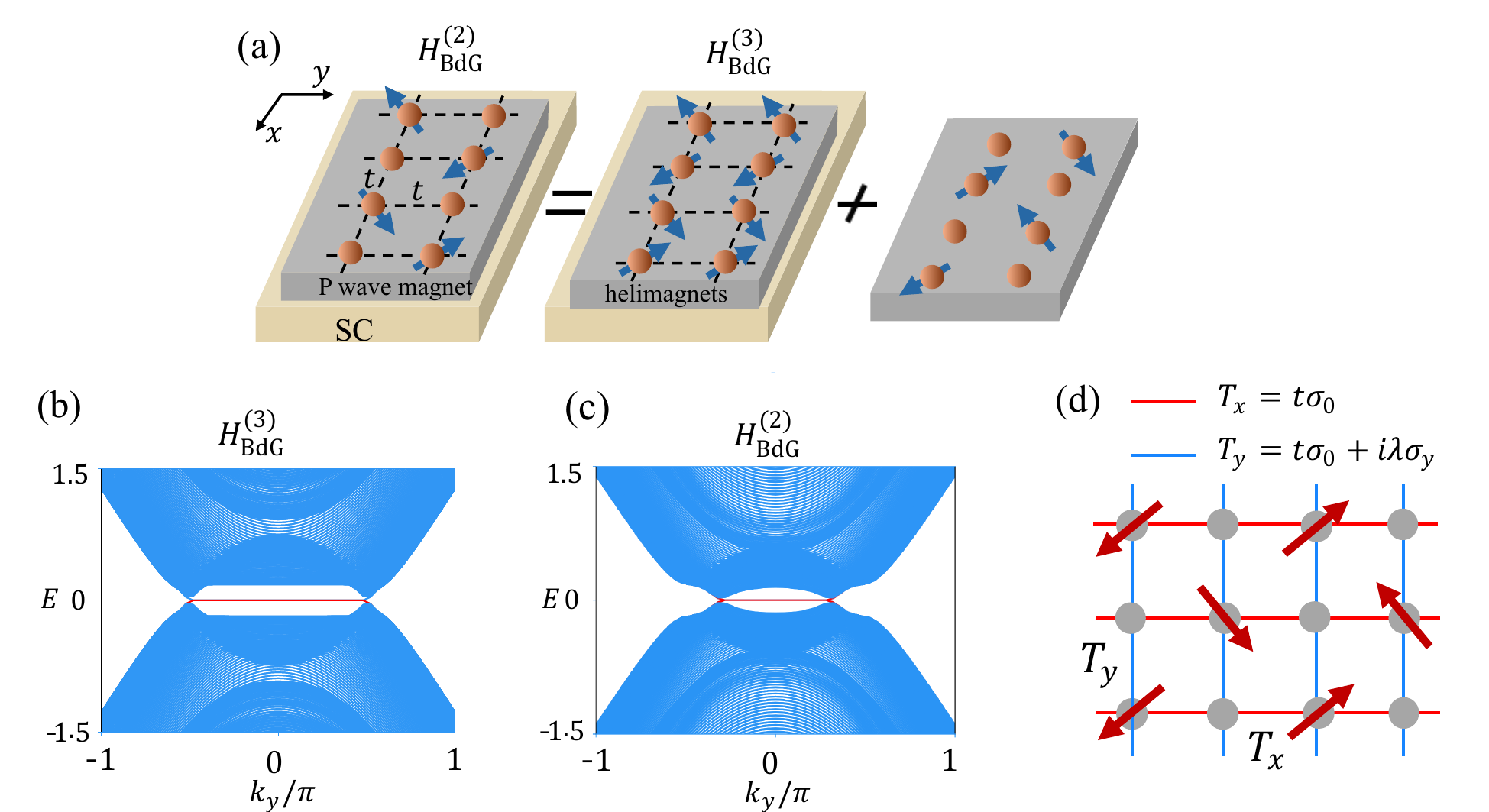}
\caption{ (a) Schematic illustration of model $H_{\text{BdG}}^{(2)}$, mapping it to stacked superconducting 1D helimagnets with additional compensated Zeeman fields. (b) and (c) Energy bands of $H_{\text{BdG}}^{(3)}$ and  $H_{\text{BdG}}^{(2)}$ in a nanowire geometry along the $y$ direction. (d) Schematic illustration of the relativistic spin-orbit coupling term  $i\lambda\sigma_y$ in $H_2$. The model parameters are taken as $t=1$, $J=0.5$, $\Delta=0.2$, $\lambda=0.1$,  and $\mu=3.5$. }
\label{Figs4}
\end{figure*}}

\subsection{B. Details of model $H_2$}
 In this section, we present the details of the model $H_2$, which is defined on a square lattice.  $H_2$ contains eight magnetic sublattices labeled A through H. The local magnetic moments are given by
\beqn
&&\bm{m}_{\text{A}} = \left(-\frac{\sqrt{2}}{2}, -\frac{\sqrt{2}}{2}, 0\right), 
\bm{m}_{\text{C}} = \left( \frac{\sqrt{2}}{2},  \frac{\sqrt{2}}{2}, 0\right), 
\bm{m}_{\text{F}} = \left( \frac{\sqrt{2}}{2}, -\frac{\sqrt{2}}{2}, 0\right), \nonumber \\
&&\bm{m}_{\text{H}} = \left(-\frac{\sqrt{2}}{2},  \frac{\sqrt{2}}{2}, 0\right), 
\bm{m}_{\text{B}} = \bm{m}_{\text{D}} = \bm{m}_{\text{E}} = \bm{m}_{\text{G}} = (0,0,0). \nonumber
\eeqn
The arrangement of these magnetic moments is shown in Fig.~\ref{Figs4}(a).

$H_2$ respects two effective time-reversal operations, $\tilde{\mathcal{T}}_2 = \tilde{\mathcal{T}}\tau_2$ and $\tilde{\mathcal{T}}_1 = \tilde{\mathcal{T}}\sigma_z$, together with a spin group symmetry $g_3 = \{s_y||M_x|\tau_3\}$. Here $\tau_2$ and $\tau_3$ are half-period translations along $x$ and $y$, respectively, and $M_x$ denotes a mirror reflection with respect to the $yz$-plane. These three symmetries yield the following constraints:
\begin{align}
\tilde{\mathcal{T}}_2&:\; s_z(\bm{k}) = -s_z(-\bm{k}), \qquad s_{x,y}(\bm{k}) = s_{x,y}(-\bm{k}), \nonumber\\
\tilde{\mathcal{T}}_1&:\; s_{x,y,z}(\bm{k}) = -s_{x,y,z}(-\bm{k}), \nonumber\\
g_3&:\; s_{x,z}(\bm{k}) = -s_{x,z}(M_x\bm{k}), \qquad s_y(\bm{k}) = s_y(M_x\bm{k}).
\end{align}
Consequently, we obtain $s_z(k_x,k_y) = \pm s_z(\pm k_x,\mp k_y)$ and $s_{x,y}(\bm{k}) = 0$. By expanding $s_z(\bm k)$ around the $\Gamma$ point, the leading term for describing the NSS is $k_x\sigma_z$, identifying $H_2$ as a $p$-wave magnet.

\subsection{C. TSCs based on models $H_{1,2}$}
\subsection{C1. Majorana flat bands}
By introducing $s$-wave superconducting pairing into models $H_{1}$ and $H_{2}$, we construct the corresponding Bogoliubov–de Gennes (BdG) Hamiltonians, denoted as $H_{\text{BdG}}^{(1,2)}$.  The Hamiltonians \(H_{\text{BdG}}^{(1,2)}\) respect the particle–hole symmetry \(\mathcal{P} = \tau_x K\) and the effective time-reversal symmetry \(\tilde{\mathcal{T}}_2 = \mathcal{T}\tau_z\sigma_z\), giving rise to the chiral symmetry \(C = \mathcal{P} \tilde{\mathcal{T}}_2 = \tau_y\sigma_x\), which places the systems in the BDI symmetry class. In two dimensions, this chiral symmetry can stabilize a nodal topology and protect Majorana flat bands as demonstrated in the main text.

Since $H_1(\bm k)$ can be exactly mapped to the two-band model $h_0(\bm k)$, it is convenient to analyze the nodal structure of $H_{\text{BdG}}^{(1)}$ using $h_0$. For $h_0$ with an $s$-wave pairing, the BdG Hamiltonian reads
\begin{equation}
\tilde{h}_{\text{BdG}}^{(1)} = -\tilde{f}_0(\bm k) \tau_z \sigma_0 + J \tau_0 \tilde{\sigma}_x - f_z(\bm k) \tau_z \tilde{\sigma}_z + \Delta \tau_x \sigma_0,
\end{equation}
where \(\tilde{f}_0 = f_0 - \mu\).
The energy spectrum of $\tilde{h}_{\text{BdG}}^{(1)}$ is given by
\begin{equation}
E(\bm k)=\pm\sqrt{J^2+\Delta^2+\tilde{f}_0^2+f_z^2\pm 2\sqrt{J^2\Delta^2+J^2\tilde{f}_0^2+f_z^2\tilde{f}_0^2}}.
\end{equation}
{ Therefore, the gap closes if and only if
\begin{equation}
f_z(\bm{k}) = 0 \quad\text{and}\quad \sqrt{\Delta^2 + \bigl[f_0(\bm{k}) - \mu\bigr]^2} = J.
\end{equation}
The second condition can be rewritten as
\begin{equation}
\bigl[f_0(\bm{k}) - \mu\bigr]^2 = J^2 - \Delta^2,
\end{equation}
which requires $J > \Delta$; otherwise the spectrum is fully gapped for any chemical potential. Assuming $J > \Delta$, the node points exist precisely at those momenta satisfying $f_z(\bm{k}) = 0$ and
\begin{equation}
\mu = f_0(\bm{k}) \pm \sqrt{J^2 - \Delta^2}.
\end{equation}
The condition $f_z(\bm{k}) = 0$ defines the high-symmetry lines in the Brillouin zone (e.g., the $\Gamma$--$\tilde{M}$ lines for the triangular lattice). Along these lines, $f_0(\bm{k})$ reaches its maximum value $3t$ at the $\Gamma$ point and its minimum value $-t$ at the $\tilde{M}$ point. Thus, the range of $f_0(\bm{k})$ is $[-t,\, 3t]$, and the topological nodal superconducting state---characterized by the existence of gapless Majorana points---is supported when the chemical potential lies in the range
\begin{equation}
\,\mu \in \left[-t - \sqrt{J^2 - \Delta^2},\; 3t + \sqrt{J^2 - \Delta^2}\,\right]\quad\text{with}\quad J > \Delta.\,
\label{MZMC}
\end{equation}
Outside this range, the BdG spectrum is fully gapped and the system is a trivial superconductor. By contrast, the topological region for $\mu$ in model $H_{\mathrm{BdG}}^{(2)}$ is determined numerically, as the square-lattice $p$-wave magnet does not admit a simple analytical expression for the nodal condition.}

\begin{figure*}
\centering
\includegraphics[width=\textwidth]{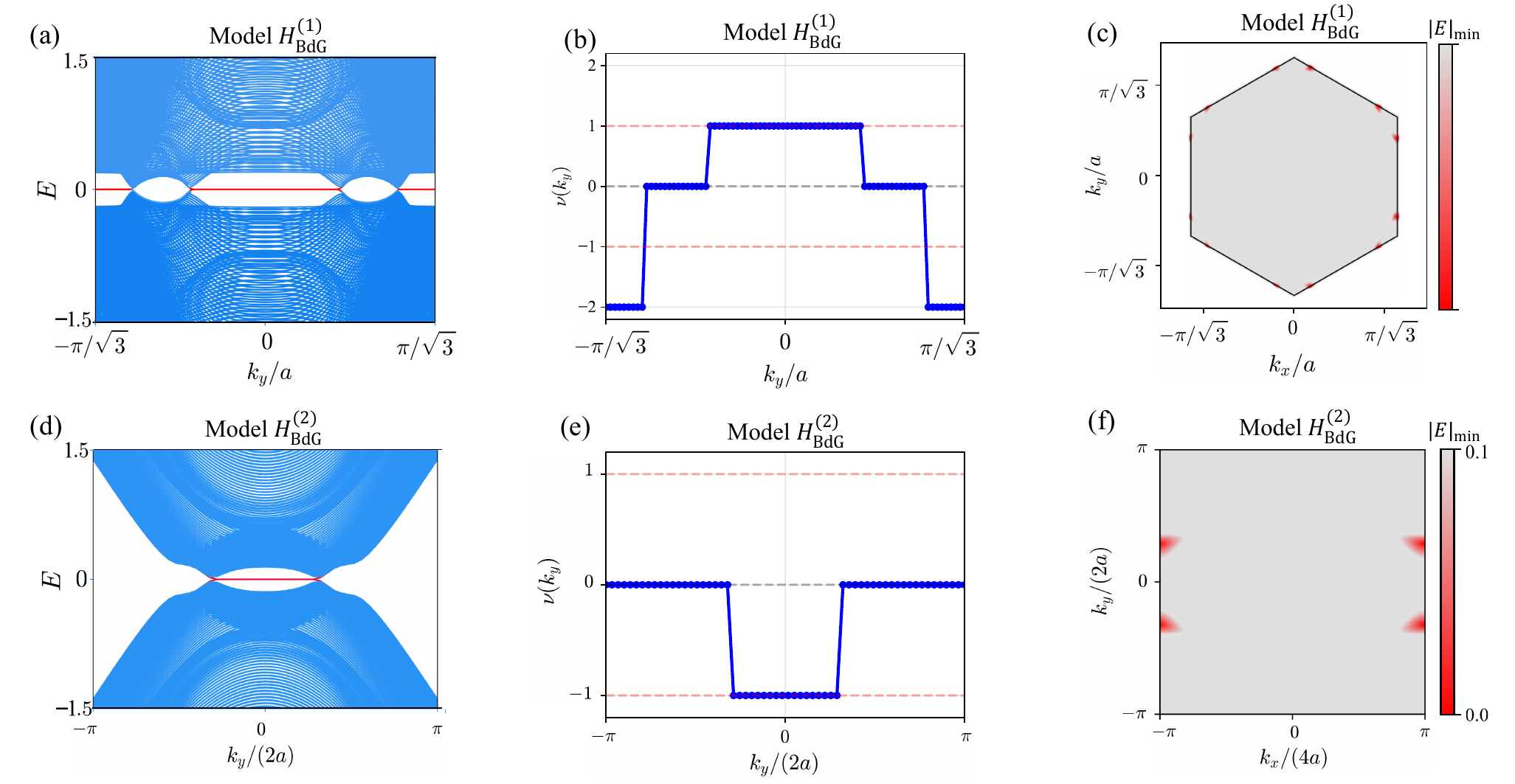}
\caption{{(a) Energy spectrum of $H_{\mathrm{BdG}}^{(1)}$ in a nanowire geometry with open boundary conditions along $x$ and periodic boundary conditions along $y$, exhibiting Majorana flat bands (red lines near zero energy). (b) Momentum-dependent winding number $\nu(k_y)$ for $H_{\mathrm{BdG}}^{(1)}$, whose quantized plateaus provide a one-to-one characterization of the Majorana flat bands. (c) Color map of the minimum excitation gap $|E|_{\min}$ in the magnetic Brillouin zone for model $H_1$, showing six nodal points. (d)--(f) Corresponding results for $H_{\mathrm{BdG}}^{(2)}$: (d) energy spectrum with Majorana flat bands, (e) winding number $\nu(k_y)$, and (f) $|E|_{\min}$ color map showing four nodal points. The common parameters are $t=1$, $J=0.5$, and $\Delta=0.2$ (with $\mu=1$ for $H_{\mathrm{BdG}}^{(1)}$ and $\mu=3.5$ for $H_{\mathrm{BdG}}^{(2)}$).}}
\label{fig:majorana}
\end{figure*}

{
The topological protection of the Majorana flat bands is characterized by the momentum-dependent winding number $\nu(k_y)$. In the eigenbasis of the chiral symmetry $C$, the BdG Hamiltonian can be brought into an off-diagonal block form
\begin{equation}
\tilde{H}_{\mathrm{BdG}}^{(i)}(\bm{k}) =
\begin{pmatrix}
0 & A(\bm{k})\\[4pt]
A^{\dagger}(\bm{k}) & 0
\end{pmatrix},
\qquad
\tilde{H}_{\mathrm{BdG}}^{(i)} = U H_{\mathrm{BdG}}^{(i)} U^{\dagger},
\end{equation}
where the unitary matrix $U$ diagonalizes $C$ such that $U C U^{\dagger}$ is block-diagonal with eigenvalues $+1$ and $-1$ in the respective subspaces. The winding number is then defined as the integral of the $U(1)$ phase of the determinant of the off-diagonal block $A(\bm{k})$:
\begin{equation}
\nu(k_y) = \frac{1}{2\pi}\int_{-\pi}^{\pi}\!dk_x\, \partial_{k_x}\text{ln} z(\bm{k}),
\qquad
z(\bm{k}) = \frac{\det A(\bm{k})}{|\det A(\bm{k})|}.
\end{equation}
As shown in Fig.~\ref{fig:majorana}(b), $\nu(k_y)$ takes quantized integer values and changes discontinuously at the gap-closing momenta, providing a one-to-one characterization of the Majorana flat bands shown in Fig.~\ref{fig:majorana}(a).
The bulk nodal structure is further visualized in Fig.~\ref{fig:majorana}(c) through the color map of the minimum excitation gap $|E|_{\min}$ in the magnetic Brillouin zone, showing six nodal points. The corresponding results for $H_{\mathrm{BdG}}^{(2)}$  introduced in the main text are shown in Figs.~\ref{fig:majorana}(d)--\ref{fig:majorana}(f), where four nodal points appear.}

{
To clarify the relation between the two-band model and the full six-band model, we plot in Fig.~\ref{fig:nodefolding}(a) the minimum gap $|E|_{\min}$ of the BdG model $H_{\mathrm{BdG}}^{(0)}$ (i.e., the BdG Hamiltonian constructed from $h_0(\bm k)$) in the original lattice Brillouin zone, which exhibits six node points located along the $\Gamma$--$\tilde{M}_{1,2,3}$ lines. Figure~\ref{fig:nodefolding}(d) shows the folding of these node points into the magnetic Brillouin zone of the model $H_{\mathrm{BdG}}^{(1)}$, demonstrating a precise one-to-one correspondence.}

\begin{figure*}
\centering
\includegraphics[width=0.75\textwidth]{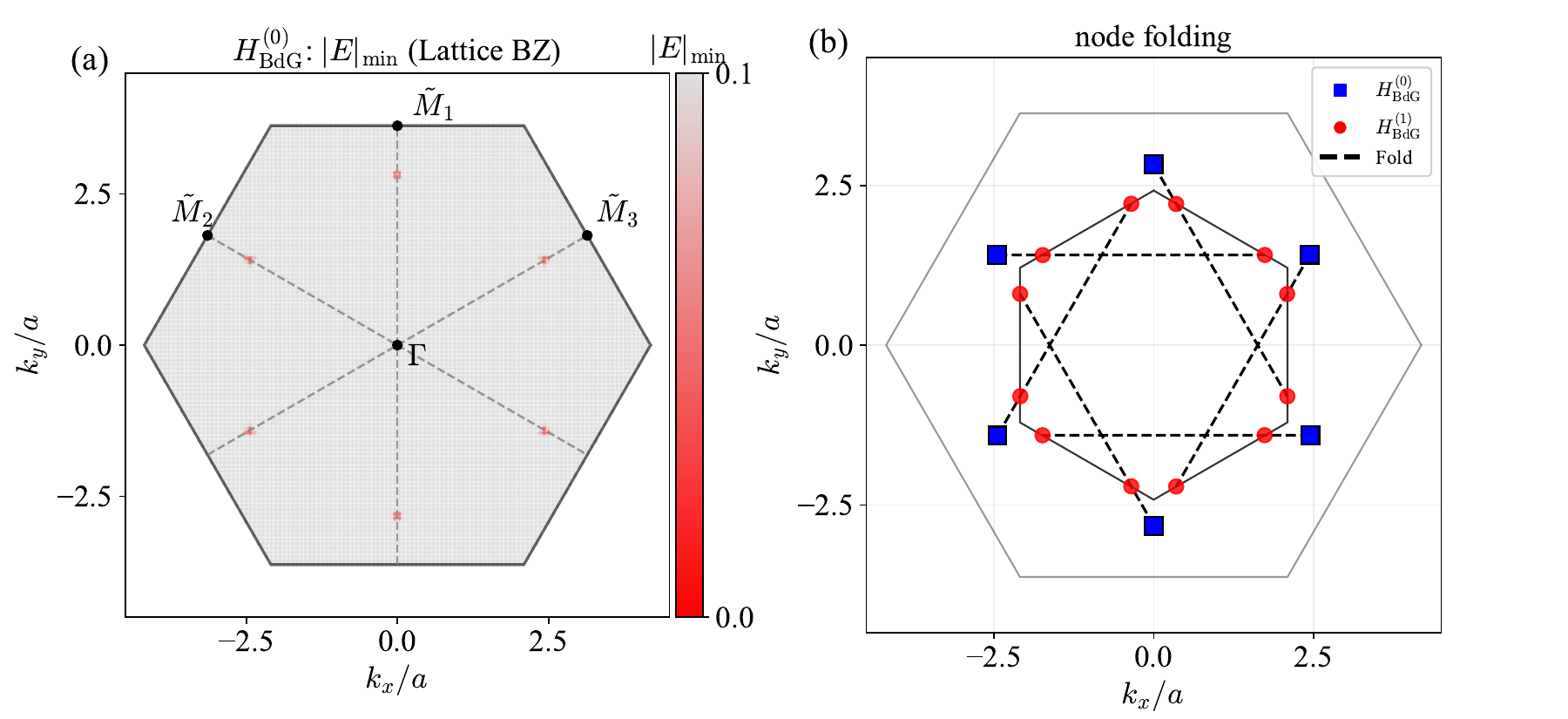}
\caption{{(a) Color map of the minimum absolute energy $|E|_{\min}$ of the model $H_{\mathrm{BdG}}^{(0)}$ in the original lattice Brillouin zone, showing six node points located along the high-symmetry lines. (b) Mapping of nodal points between the model $H_{\mathrm{BdG}}^{(0)}$ (blue squares) and the model $H_{\mathrm{BdG}}^{(1)}$ (red circles), illustrating the band-folding relation. The parameters are $t=1$, $J=0.5$, $\Delta=0.2$, and $\mu=1$.}}
\label{fig:nodefolding}
\end{figure*}

As explained in Fig.~\ref{Figs4}(a), the model $H_{\text{BdG}}^{(2)}$ can be decomposed into a stack of superconducting helimagnets with additional compensated Zeeman fields. A stack of superconducting helimagnets hosts the nodal topology featuring Majorana flat bands, as shown in Fig.~\ref{Figs4}(b). By comparing Fig.~\ref{Figs4}(c) with Fig.~\ref{Figs4}(b), one sees that the extra Zeeman
fields merely shift the positions of the bulk nodes while the flat bands remain.

\begin{figure}[t]
\centering
\includegraphics[width=5in]{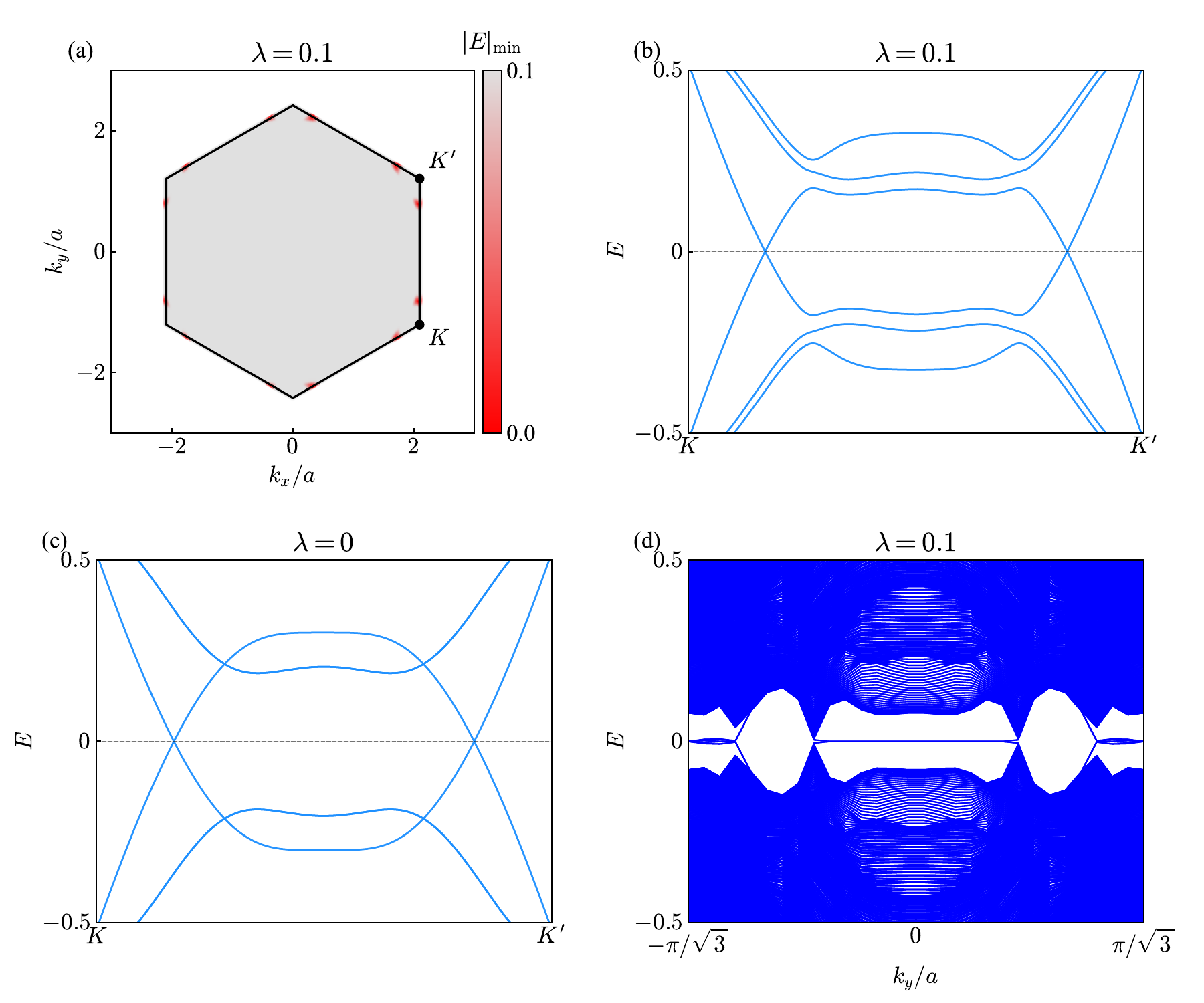}
\caption{
{(a) Color map of the minimum excitation gap $|E|_{\min}$ in the magnetic Brillouin zone for model $H_{\mathrm{BdG}}^{(1)}$ with weak SOC ($\lambda=0.1$), showing that the bulk nodal points remain gapless. (b) Energy spectrum along the $K$--$K'$ path with $\lambda=0.1$. (c) Same as (b) but without SOC ($\lambda=0$) for comparison. (d) Energy spectrum under open boundary conditions along the $x$ direction with system size $L_x=500$ and $\lambda=0.1$, showing that the zero modes remain visible. The other parameters are $t=1$, $J=0.5$, $\Delta=0.2$, and $\mu=1$.}}
\label{figR8}
\end{figure}

\subsection{C2. Spin-orbit coupling effects}

Based on the model $H_{2}$, we further include a relativistic spin-orbital coupling (SOC) term $i\lambda \sigma_y$ along the $y$ direction, as schematically depicted in Fig.~\ref{Figs4}(d). In the presence of this term, the original  Majorana flat bands shown in Fig.~\ref{Figs4}(c) acquire dispersion and become propagating edge states, while the bulk nodal structure remains intact. This result is displayed in Fig.~3(e) of the main text, where the SOC strength is taken as $\lambda = 0.1$.

{
In model $H_1$, the SOC term  spoils the analytic solvability  of the Hamiltonian, and the system can no longer be mapped to the effective two-band model $h_0(\bm{k})$. We therefore resort to numerical studies. Specifically, we consider the Hamiltonian
\beqn
H_1'(\bm{k}) =
\begin{pmatrix}
-\frac{J}{2}(\sqrt{3}\sigma_x+\sigma_y) & f(\bm{k})M & f^{*}(\bm{k})M^{*} \\[4pt]
f^{*}(\bm{k})M^{*} & \frac{J}{2}(\sqrt{3}\sigma_x-\sigma_y) & f(\bm{k})M \\[4pt]
f(\bm{k})M & f^{*}(\bm{k})M^{*} & J\sigma_y
\end{pmatrix},
\eeqn
where $M = t\sigma_0 + i\lambda\sigma_y$ with $\lambda$ being the SOC strength. The corresponding BdG Hamiltonian is constructed as
\beqn
H_{\mathrm{BdG}}'(\bm{k}) =
\begin{pmatrix}
H_1'(\bm{k})-\mu & i\Delta\sigma_y\\
-i\Delta^{*}\sigma_y & -H_1'^{*}(-\bm{k})+\mu
\end{pmatrix}.
\eeqn}
{
Our numerical results show that a weak SOC ($\lambda=0.1$) leaves the bulk nodal structure intact. The node points remain gapless, as evidenced by the $|E|_{\min}$ colormap of $H_{\mathrm{BdG}}'(\bm{k})$ in the magnetic Brillouin zone [Fig.~\ref{figR8}(a)]. Along the $K$--$K'$ path, the energy spectrum with SOC [Fig.~\ref{figR8}(b)] remains nearly unchanged compared to the $\lambda=0$ case [Fig.~\ref{figR8}(c)], and the zero-energy crossings associated with the bulk nodes persist. Under open boundary conditions, the zero modes remain visible in the spectrum [Fig.~\ref{figR8}(d)]. However, the Majorana flat bands are no longer protected by symmetry, because the SOC term explicitly breaks the chiral symmetry $C$. These results indicate that the effect of SOC on Majorana flat bands is model-dependent: in $H_1$ the SOC destroys the symmetry protection while preserving the nodal bulk, whereas in $H_2$ it drives the flat bands into dispersive edge states.}

\subsection{D. Engineering TSCs Based on 1D OPMs}
OPMs are magnetic systems characterized by a NSS that is odd under spatial inversion, satisfying \(\bm{S}(\bm{k}) = -\bm{S}(-\bm{k})\), where \(\bm{S}(\bm{k}) = (s_x(\bm{k}), s_y(\bm{k}), s_z(\bm{k}))\). Based on the dimensionality of \(\bm{S}(\bm{k})\), OPMs are classified into three distinct types \cite{2025luospin}. Type-I OPMs exhibit a collinear spin texture, for example, \(s_z(\bm{k}) = -s_z(-\bm{k})\) with \(s_x(\bm{k}) = s_y(\bm{k}) = 0\). Type-II OPMs possess a coplanar spin texture, where the spins are confined to a plane, for example \(s_z(\bm{k}) = 0\) and \(s_{x,y}(\bm{k}) = -s_{x,y}(-\bm{k})\). Type-III OPMs are characterized by a non-coplanar spin texture, with all three spin components being nonzero and each obeying \(s_\alpha(\bm{k}) = -s_\alpha(-\bm{k})\) for \(\alpha = x, y, z\). Therefore, the previously studied 2D models $H_{1,2}$ belong to type-I OPMs.
In this section, we further use three 1D models, which belong to type-I, type-II, and type-III OPMs, respectively, to engineer TSCs.

\subsection{D1. Model Hamiltonians and TSCs}
 Type-I OPMs can be realized in coplanar magnetic orders respecting the spin-group symmetry \(g_1=\{U_z(\theta)|\tau\}\), such as the 1D helimagnet described by model \(\mathcal{H}_1\), as schematically illustrated in Fig.~\ref{Figs1}(a). Type-II OPMs can be realized in non-coplanar magnetic orders that simultaneously possess the symmetries \(\mathcal{T}\tau\) and \(\{U_z(\pi)||P\}\) (with \(P\) denoting spatial inversion), as implemented in model \(\mathcal{H}_2\), schematically illustrated in Fig.~\ref{Figs1}(b). Type-III OPMs are realized in non-coplanar magnetic orders respecting the \(\mathcal{T}\tau\) symmetry, exemplified by model \(\mathcal{H}_3\), as schematically illustrated in Fig.~\ref{Figs1}(c).

\begin{figure*}
\centering
\includegraphics[width=5.7in]{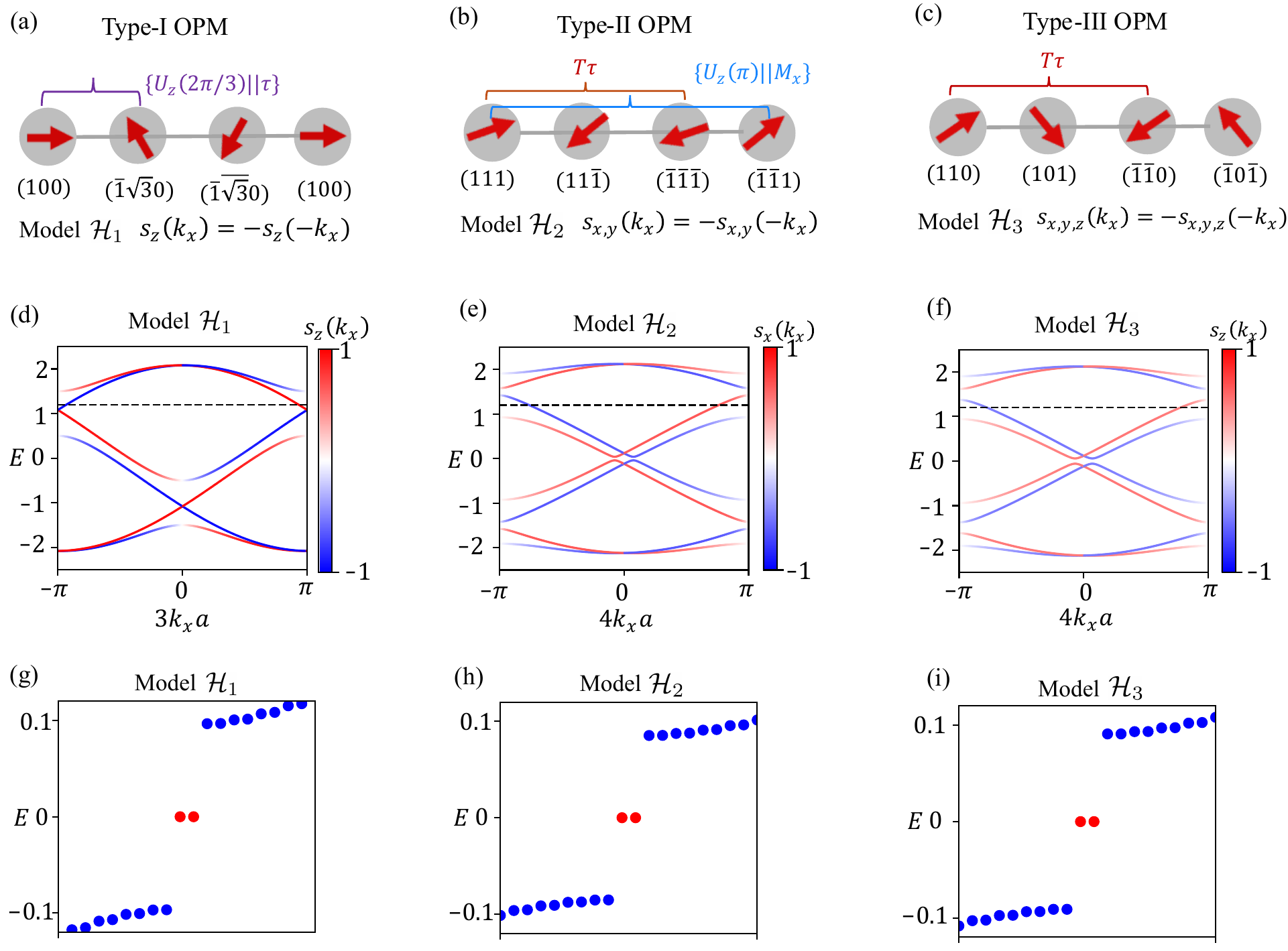}
\caption{  (a), (b), and (c) Schematic illustration of the theoretical models $\mathcal{H}_1$, $\mathcal{H}_2$, and $\mathcal{H}_3$, realizing type-I, type-II, and type-III OPMs, respectively. The numbers $(xyz)$ beneath the magnetic atoms denote the directional vectors of the local magnetic moments. Solid lines indicate that different magnetic sublattices are related by specific symmetries. 
 (d), (e), and (f) Bulk energy bands for models $\mathcal{H}_1$, $\mathcal{H}_2$, and $\mathcal{H}_3$, respectively. 
(g), (h), and (i) Eigenvalues near zero energy for models $\mathcal{H}_{\text{BdG}}^{(i)}$ under the open boundary conditions. The adopted chemical potential is marked by black dashed lines in (d)--(f)),  lying within the magnetic gap. The model parameters are taken as $t=1$, $J=0.5$, and $\Delta=0.1$. }
\label{Figs1}
\end{figure*}

The model Hamiltonians \(\mathcal{H}_{1,2,3}\) can be expressed in a unified form:
\begin{equation}
\mathcal{H}_{n} = \sum_{\langle \bm{r}\bm{r}'\rangle,\sigma} t \, c^\dagger_{\bm{r}\sigma} c_{\bm{r}'\sigma}
+ J \sum_{\bm{r},\sigma\sigma'} \bm{m}_{n}(\bm{r}) \cdot c^\dagger_{\bm{r}\sigma} \bm{\sigma}_{\sigma\sigma'} c_{\bm{r}\sigma'}, \quad n=1,2,3
\end{equation}
For $\mathcal{H}_1$, the local magnetic moment is expressed as \(\bm{m}_1(\bm r) = \big( \cos(\bm {Q}_1\cdot \bm r ),\, \sin(\bm {Q}_1\cdot \bm r),\, 0 \big)\),   where \(|\bm Q_1| = 2\pi / \mathcal{L}\). In Fig.~\ref{Figs1}(a), we take \(\mathcal{L}=3\) and $\mathcal{H}_1$ realizes a $p$-wave magnet. When \(\mathcal{L}=1\) and \(\mathcal{L}=2\), \(\mathcal{H}_1\) describes a conventional ferromagnet and an antiferromagnet, respectively. Models \(\mathcal{H}_2\) and \(\mathcal{H}_3\) have four magnetic sublattices. Their local moments are given by:
For \(\mathcal{H}_2\) (sublattices A, B, C, D):
\[
\bm{m}_2^{\mathrm{A}} = \left(\frac{1}{\sqrt{3}},\frac{1}{\sqrt{3}},\frac{1}{\sqrt{3}}\right),\quad
\bm{m}_2^{\mathrm{B}} = \left(\frac{1}{\sqrt{3}},\frac{1}{\sqrt{3}},-\frac{1}{\sqrt{3}}\right),\quad
\bm{m}_2^{\mathrm{C}} = -\left(\frac{1}{\sqrt{3}},\frac{1}{\sqrt{3}},\frac{1}{\sqrt{3}}\right),\quad
\bm{m}_2^{\mathrm{D}} = \left(-\frac{1}{\sqrt{3}},-\frac{1}{\sqrt{3}},\frac{1}{\sqrt{3}}\right).
\]
For \(\mathcal{H}_3\) (sublattices A, B, C, D):
\[
\bm{m}_3^{\mathrm{A}} = \left(\frac{1}{\sqrt{2}},\frac{1}{\sqrt{2}},0\right),\quad
\bm{m}_3^{\mathrm{B}} = \left(\frac{1}{\sqrt{2}},0,\frac{1}{\sqrt{2}}\right),\quad
\bm{m}_3^{\mathrm{C}} = \left(-\frac{1}{\sqrt{2}},-\frac{1}{\sqrt{2}},0\right),\quad
\bm{m}_3^{\mathrm{D}} = \left(-\frac{1}{\sqrt{2}},0,-\frac{1}{\sqrt{2}}\right).
\]
Under the corresponding symmetry constraints, the spin textures of the models $\mathcal{H}_{1,2,3} $ satisfy \cite{2025luospin}:
\beqn
    &&\mathcal{H}_1: s_z(k_x)=-s_z(-k_x), \quad s_{x,y}(k_x)=0;\nonumber\\
    &&\mathcal{H}_2: s_{x,y}(k_x)=-s_{x,y}(-k_x), s_z(k_x)=0;\nonumber\\
    &&\mathcal{H}_3: s_{x,y,z}(k_x)=-s_{x,y,z}(-k_x) .
\eeqn
These relations are fully consistent with the numerical results. The corresponding bulk energy bands of $\mathcal{H}_{1,2,3}$ are shown in Figs.~\ref{Figs1}(d)-\ref{Figs1}(f).

A key feature in the band structures of \(\mathcal{H}_1\), \(\mathcal{H}_2\), and \(\mathcal{H}_3\) is that for some bands, the Kramers degeneracy at time-reversal invariant momenta is lifted. Specifically, certain energy bands of \(\mathcal{H}_1\) exhibit no Kramers degeneracy at either \(k_x = 0\) or \(k_x = \pi\), while for \(\mathcal{H}_2\) and \(\mathcal{H}_3\), energy bands retain degeneracy only at \(k_x = 0\). This selective lifting of degeneracy in $\mathcal{H}_{2,3}$  follows from the symmetry \(\tilde{\mathcal{T}}_1 = \mathcal{T}\tau\), with $\tilde{\mathcal{T}}_1^2=-1$ at $k_x=0$. The systematic absence of Kramers pairs at  time-reversal-invariant momenta in these bands signals the presence of a hidden Zeeman field in OPMs, which opens a magnetic energy gap. With this gap opened, introducing \(s\)-wave superconducting pairing drives the systems into a 1D topological superconducting phase featuring Majorana end modes, as shown in Figs.~\ref{Figs1}(g)--\ref{Figs1}(i).

\subsection{D2. Band structures of the 1D helimagnets}
To analytically reveal the band structure of \(\mathcal{H}_1\) for $\mathcal{L}=3$, we perform a local unitary transformation that aligns the magnetic moments along the \(x\)-axis:
\begin{equation}
\tilde{\mathcal{H}}_1 = \mathcal{U}_1 \mathcal{H}_1 \mathcal{U}_1^\dagger, \quad \text{with} \quad \mathcal{U}_1(\bm r) = e^{-i \bm Q_1 \cdot \bm r \sigma_z / 2}.
\end{equation}
The transformed Hamiltonian becomes
\begin{equation}
\tilde{\mathcal{H}}_1 = \sum_{\langle \bm r\bm r'\rangle,\sigma\sigma'} \left( \tilde{t}_{\sigma\sigma'} \, \tilde{c}_{\bm r\sigma}^\dagger \tilde{c}_{\bm r'\sigma'} + \text{h.c.} \right)
+ J \sum_{\bm r,\sigma\sigma'} \tilde{c}_{\bm r\sigma}^\dagger (\tilde{\sigma}_x)_{\sigma\sigma'} \tilde{c}_{\bm r\sigma'},
\end{equation}
where \(\tilde{c}_{\bm r\sigma} =\sum_{\sigma'} \mathcal{U}_{1,\sigma\sigma'}r(\bm r) c_{\bm r\sigma'}\), and \(\tilde{t}_{\sigma\sigma'} = t[\mathcal{U}_1(\bm r)\mathcal{U}_1^{\dagger}(\bm r')]_{\sigma\sigma'}\). In the new basis, the hopping \(\tilde{t}_{\sigma\sigma'}\) is $te^{-i\pi \tilde{\sigma}_z/3}$, which reveals the original of NSS. The second term reveals the  origin of hidden Zeeman field with a strength of \(J\).

It is important to note that \(\tilde{\mathcal{H}}_{1}\) restores the translation symmetry of the underlying lattice, which simplifies the analysis and allows us to derive an effective two-band description. In the rotated frame, the periodic boundary conditions are modified to anti-periodic ones \cite{NadjPerge2013} and we obtain the Bloch Hamiltonian \cite{Martin2012}
\begin{equation}
\tilde{\mathcal{H}}_1(k) = J\sigma_x - t\cos k_x\,\sigma_0 - \sqrt{3}t\sin k_x\,\sigma_z.
\end{equation}
The energy bands of $\mathcal{H}_1(\bm k)$ can be obtained by folding  those of $\tilde{\mathcal{H}}_1(k)$.

\begin{figure*}
\centering
\includegraphics[width=4.in]{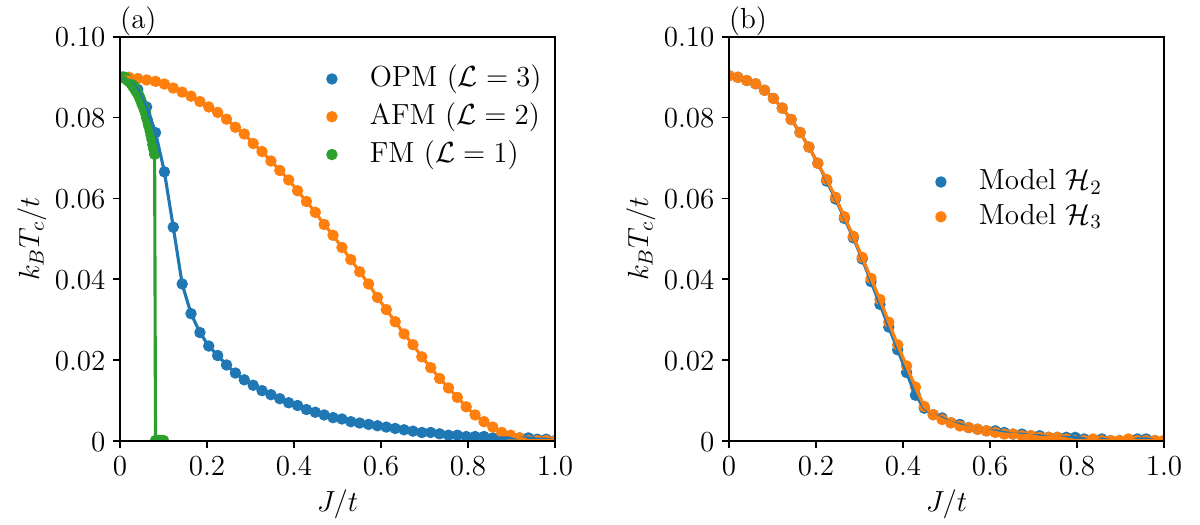}
\caption{ (a) Self-consistently calculated transition temperature $T_c$ versus $J/t$ for model $\mathcal{H}_{\text{BdG}}^{(1)}$, showing distinct magnetic phases: OPM with $\mathcal{L}=3$, antiferromagnetic (AFM) with $\mathcal{L}=2$, and ferromagnetic (FM) with $\mathcal{L}=1$. (b) Self-consistently calculated transition temperature for models $\mathcal{H}_{\text{BdG}}^{(2)}$ and $\mathcal{H}_{\text{BdG}}^{(3)}$. The common model parameters are $t=1$, $J=0.5$, $U=1.5$, and $\mu=-1$. }
\label{Figs1d}
\end{figure*}

\subsection{E. Robustness of superconductivity}

\subsection{E1. Compatibility between different types of OPMs and superconductivity}
In the main text, we have shown that superconducting pairing is compatible with type-I OPMs with collinear spin textures. In this subsection, we further demonstrate that this is a general characteristic, regardless of whether the spin textures of the OPMs are collinear, coplanar, or non-coplanar.

To describe the superconducting state, we introduce an attractive on-site interaction \(-U\;(U>0)\) that mediates an \(s\)-wave pairing. The total lattice Hamiltonian is
\begin{equation}
\mathcal{H}_{\text{SC}}^{(i)} = \mathcal{H}_i - U \sum_{\bm r} c_{\bm r\uparrow}^{\dagger}c_{\bm r\downarrow}^{\dagger}c_{\bm r\downarrow}c_{\bm r\uparrow},
\end{equation}
where \(i=1,2,3\) labels the model. Within the mean-field approximation, the uniform superconducting order parameter is defined as \(\Delta = U\langle c_{\bm r\downarrow}c_{\bm r\uparrow}\rangle\). The corresponding BdG Hamiltonian in momentum space reads
\begin{equation}
\mathcal{H}_{\text{BdG}}^{(i)}(\bm k)=
\begin{pmatrix}
H_i(\bm k)-\mu & i\Delta\sigma_y\\
-i\Delta^{*}\sigma_y & -H_i^{*}(-\bm k)+\mu
\end{pmatrix},
\end{equation}
with \(\mu\) the chemical potential. The mean-field free energy density is
\begin{equation}
\mathcal{F}_s=\frac{|\Delta|^{2}}{U}-\frac{1}{2\beta V}\sum_{\bm k,n}\ln\!\Bigl(1+e^{-\beta E_{\bm k,n}}\Bigr),
\end{equation}
where \(\beta=1/(k_BT)\), \(E_{\bm k,n}\) are the eigenvalues of \(\mathcal{H}_{\text{BdG}}^{(i)}(\bm k)\), and $V$ represents the volume of the system. The critical temperature \(T_c\) is obtained from the linearized gap equation
\begin{equation}
\left.\frac{\partial^{2}\mathcal{F}_s(T,\Delta)}{\partial\Delta^{2}}\right|_{\Delta=0}=0,
\end{equation}
which we solve numerically to obtain \(T_c\).

Figure.~\ref{Figs1d}(a) shows the normalized critical temperature $T_c$ as a function of $J/t$ for model $\mathcal{H}_{\text{BdG}}^{(1)}$ with different magnetic periods \(\mathcal{L}\). For the ferromagnetic case (\(\mathcal{L}=1\)), $T_c$ drops sharply because the uniform exchange field fully polarizes the spins and strongly suppresses spin‑singlet pairing. In the antiferromagnetic case (\(\mathcal{L}=2\)), superconductivity persists up to a larger critical coupling \(J_c/t \simeq 0.9\). This enhanced stability stems from the spin‑degenerate nature of the antiferromagnetic state, which is intrinsically compatible with singlet pairing. For \(\mathcal{L}=3\), where \(\mathcal{H}_1\) realizes a type‑I OPM, superconductivity survives up to \(J_c/t \simeq 0.8\). Although the net magnetization of this OPM is zero, it exhibits substantial NSS, which reduces the protection compared to a simple antiferromagnet. Nevertheless, the critical \(J_c\) remains significantly higher than that of the ferromagnet, confirming that the odd-parity NSS inherent in the OPM robustly protects the superconducting state from the exchange field.

Notably, the character of the superconducting transition also differs among the three cases. For the ferromagnet (\(\mathcal{L}=1\)), the sharp drop in $T_c$ is consistent with a first-order phase transition. In contrast, for both the conventional antiferromagnet (\(\mathcal{L}=2\)) and the OPM (\(\mathcal{L}=3\)), $T_c$ decreases continuously to zero, indicative of a second-order transition. This continuous behavior reflects the protective role of spin degeneracy (for the antiferromagnet) and the odd-parity NSS (for the OPM), which stabilize the pairing against the in-plane hidden Zeeman field \cite{Yingmingxie2020,Sukhachov2025,2025ziting}. Similar trends are observed for models $\mathcal{H}_{\text{BdG}}^{(2)}$ and $\mathcal{H}_{\text{BdG}}^{(3)}$, as shown in Fig.~\ref{Figs1d}(b), despite the fact that their normal-state spin textures are coplanar and non-coplanar, respectively. In both cases, superconductivity coexists with the magnetic order over a wide range of $J$, and $T_c$ vanishes continuously, characteristic of a second-order transition.

\begin{figure*}
\centering
\includegraphics[width=0.85\textwidth]{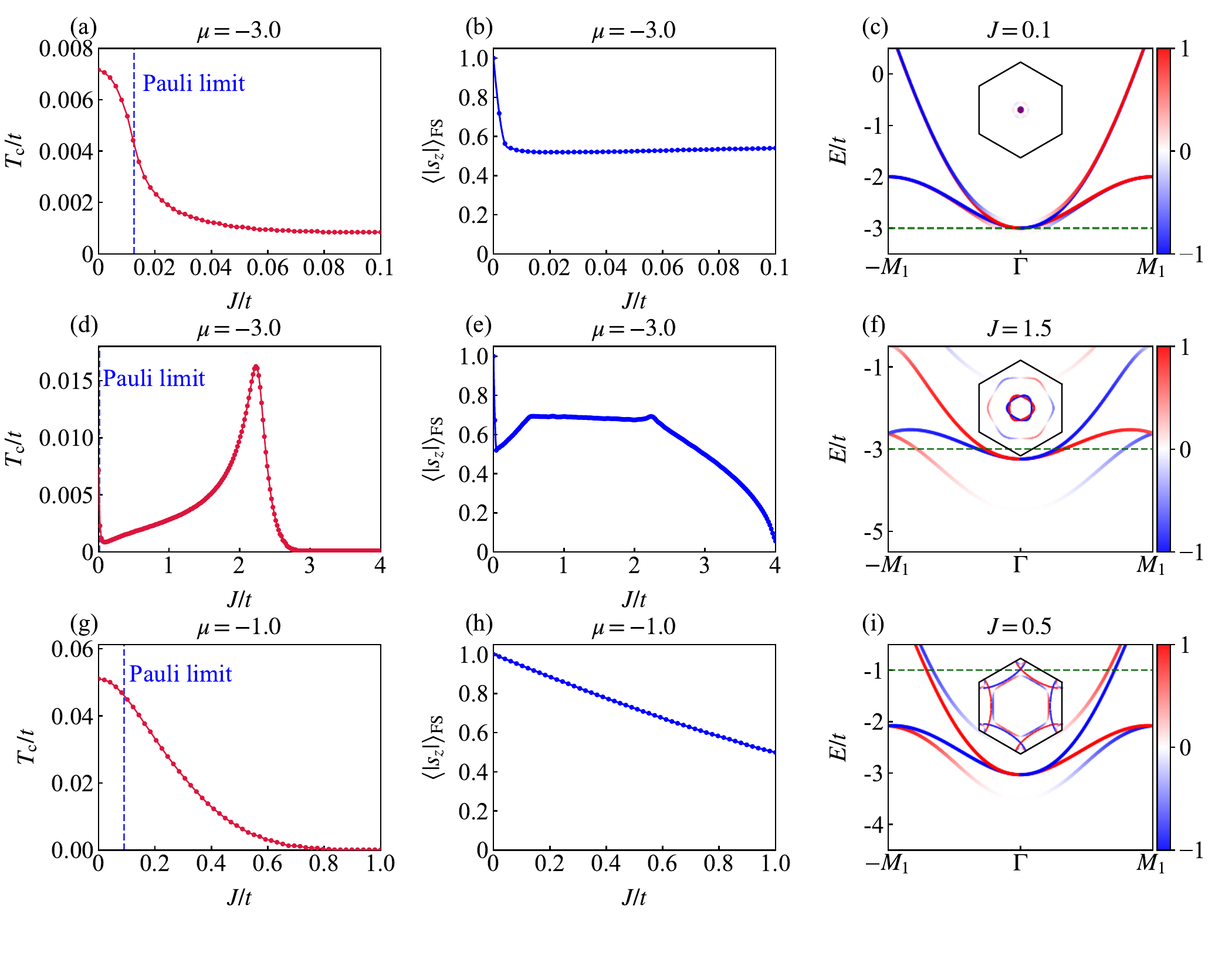}
\caption{{(a) Self-consistently calculated $T_c$ versus $J/t$ for $\mu=-3.0$ and $J/t\in[0,0.1]$. (b) Corresponding Fermi-surface averaged out-of-plane spin polarization $\langle |s_z| \rangle_{\rm FS}$. (c) Energy bands of model $H_1$ along the $M_1$--$\Gamma$--$M_1$ path at $\mu=-3.0$ and $J=0.1$; the inset plots the isoenergy surface. (d) Same as (a) but for $J/t\in[0,4]$, showing the non-monotonic evolution of $T_c$. (e) Same as (b) but for $J/t\in[0,4]$. (f) Energy bands at $\mu=-3.0$ and $J=1.5$; the inset shows the isoenergy surface. (g) $T_c$ versus $J/t$ for $\mu=-1.0$. (h) Corresponding $\langle |s_z| \rangle_{\rm FS}$ for $\mu=-1.0$. (i) Energy bands at $\mu=-1.0$ and $J=0.5$; the inset plots the isoenergy surface. The parameters are $t=1$ and $U=1.5$.}}
\label{fig:tc}
\end{figure*}

\subsection{E2. Quantifying Ising-like protection via the Fermi-surface spin polarization}

In Sec.~E1, we have demonstrated that superconductivity is compatible with OPMs regardless of whether their spin textures are collinear, coplanar, or non-coplanar. The robustness of superconductivity originates from the large time-reversal-preserving NSS, which is enforced by the spin group symmetry. To establish a quantitative link between this NSS and the superconducting critical temperature, we define the Fermi-surface averaged out-of-plane spin polarization
\begin{equation}
\langle |s_z| \rangle_{\rm FS} \equiv \frac{1}{N_{\rm FS}}\sum_{\bm{k}\in{\rm FS}} |s_z(\bm{k})|,
\end{equation}
where the sum runs over all $\bm{k}$ points on the Fermi surface. This quantity serves as a direct measure of the Ising-like protection at the Fermi level: a larger $\langle |s_z| \rangle_{\rm FS}$ implies a stronger spin-locking effect that pins the spins out of plane, thereby suppressing the pair-breaking effect of the in-plane hidden Zeeman field.

To verify this quantitative correlation, we perform self-consistent calculations for model $\mathcal{H}_{\text{BdG}}^{(1)}$ at two representative chemical potentials: $\mu=-3.0$ (the lower band edge at the $\Gamma$ point when $J=0$), and $\mu=-1.0$, which was studied in the main text. For $\mu=-3.0$, in the regime $J/t\in[0,0.1]$, $T_c$ initially drops rapidly and then saturates to a small but finite value [Fig.~\ref{fig:tc}(a)]. Correspondingly, $\langle |s_z| \rangle_{\rm FS}$ quickly decreases and then plateaus at a nearly constant value [Fig.~\ref{fig:tc}(b)]. In this regime, the Fermi level primarily intersects the lowest band [Fig.~\ref{fig:tc}(c)], which carries a finite but modest $s_z(\bm{k})$. As $J$ is further increased beyond $0.1t$, the behavior changes qualitatively: $T_c$ first rises, reaches a maximum around $J\simeq 2.2t$, and then drops continuously to zero at a critical $J_c/t\simeq 2.53$ [Fig.~\ref{fig:tc}(d)]. Notably, this critical coupling far exceeds the Pauli limit ($J_p\approx 0.0126t$) by a factor of approximately 200. This non-monotonic evolution is accompanied by a synchronous rise and fall of $\langle |s_z| \rangle_{\rm FS}$ [Fig.~\ref{fig:tc}(e)]. The physical origin is that, with increasing $J$, the Fermi level gradually cuts through the second and third bands [Fig.~\ref{fig:tc}(f)], which carry much larger $s_z(\bm{k})$. The maximal $\langle |s_z| \rangle_{\rm FS}$ at $J\simeq 2.2t$ coincides with the peak in $T_c$, confirming that the NSS-induced Ising-like protection is directly reflected by the magnitude of the spin polarization at the Fermi surface. In contrast, for $\mu=-1.0$, both $T_c$ and $\langle |s_z| \rangle_{\rm FS}$ decrease monotonically with increasing $J$ [Figs.~\ref{fig:tc}(g) and \ref{fig:tc}(h)]. This is because the Fermi level always intersects the second and third bands [Fig.~\ref{fig:tc}(i)], whose spin polarization simply weakens as $J$ grows. 

These complementary results demonstrate that the coexistence of superconductivity with strong magnetic order in OPMs is governed by the magnitude of the NSS at the Fermi surface. A larger $\langle |s_z| \rangle_{\rm FS}$ enhances the quantitative robustness of superconductivity against the hidden Zeeman field, and the qualitative features---second-order transition and coexistence beyond the Pauli limit---are preserved regardless of the specific band-crossing configuration.

\begin{figure*}
\centering
\includegraphics[width=7in]{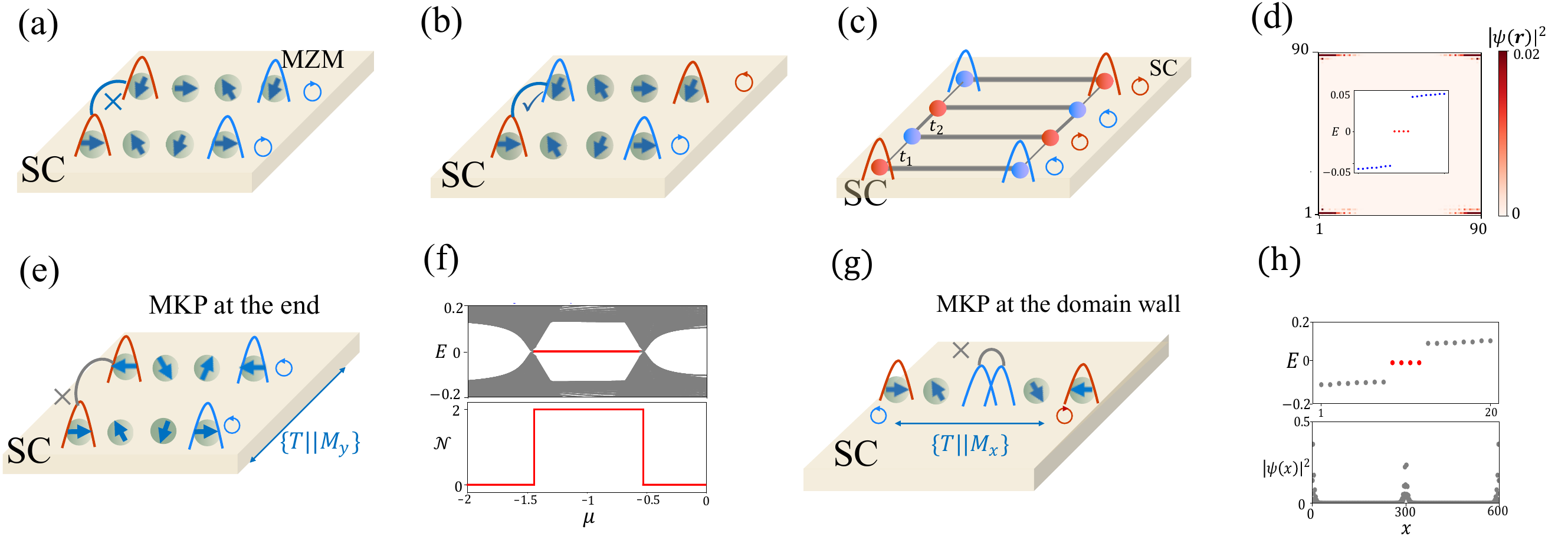}
\caption{ (a)-(b) Schematic illustration of two helimagnets with (a) identical and (b) opposite spin helicities on superconducting substrates. The two MZMs at one end cannot couple in (a) but can hybridize in (b). (c) Schematic illustration of generating Majorana corner states by a dimerized coupling of 1D TSCs formed by superconducting helimagnets. (d) Spatial distribution of Majorana corner modes in the system shown in (c); the inset displays the low-energy spectrum. (e) Illustration of  a Majorana
Kramers pair protected by the spin-group symmetry $\{T||M_y\}$. (f) Energy spectrum (upper panel) and spin-resolved Bott index $\mathcal{N}$ (lower panel) as functions of the chemical potential $\mu$ for the system in (e). (g) Majorana Kramers pairs localized at a magnetic domain wall, protected by the spin-group symmetry $\{T||M_x\}$. (h) Low-energy spectrum (upper panel) and wave functions of the four zero-energy states (lower panel, corresponding to the red dots in the upper panel) for the system in (g). Model parameters: $t=1$, $J=0.5$, $t_1=0.1$, $t_2=0.3$, $\Delta=0.2$, and $\mu=-1$. In (f) and (h), $m_z=0.1$. }
\label{Fig4}
\end{figure*}

\subsection{F. Engineering Majorana corner states and Majorana Kramers pairs}
The model $\mathcal{H}_1$ is directly relevant to the experimental platform of iron atom chains fabricated on superconducting substrates, where the helimagnetic order is stabilized by the Ruderman–Kittel–Kasuya–Yosida interaction \cite{Klinovaja2013}.  In this section, we use coupled superconducting helimagnets to realize distinct types of TSCs,  which could be realized in iron atom chains on superconductors.

\subsection{F1. Majorana end modes and corner modes}
We consider two coupled helimagnetic chains, each described by the  Hamiltonian $\mathcal{H}_{\text{BdG}}^{(1)}$ introduced before. We assume that the spin helicity \(\chi = \text{sgn}[\bm{m}(\bm r) \times \bm{m}(\bm r+\bm{a}_1)]\) of each chain can be tuned. This helicity parameter plays a crucial role in determining the topological properties of the coupled system. The helicity \(\chi\) is reversed under a \(\pi\) spin rotation about the \(y\)-axis, represented by the unitary operator \(U_y(\pi)=\sigma_y\). Because this rotation anticommutes with the chiral symmetry operator $C=\tau_x\sigma_x$,  it also reverses the chirality  of  Majorana zero modes (MZMs) present in the system, i.e., the eigenvalue of the MZMs under \(C\). Consequently, the chirality of the MZMs becomes directly linked to the spin helicity \(\chi\) of the underlying chain.

Treating the inter-chain hopping as a weak perturbation, we obtain a simple relation between the helicity configuration and the number of robust MZMs at one end of the coupled system:
\begin{equation}
N = N_{+} - N_{-} = \chi_{+} - \chi_{-},
\label{eq4}
\end{equation}
where \(N\) counts the net number of robust MZMs at one end, \(N_+\) (\(N_-\)) denotes the number of MZMs with eigenvalue \(+1\) (\(-1\)) under \(C\), and \(\chi_+\) (\(\chi_-\)) counts the number of chains with helicity \(\chi = +1\) (\(\chi = -1\)). This relation implies that two chains with the same helicity contribute two MZMs of the same chirality, which cannot hybridize and thus remain degenerate, as schematically illustrated in Fig.~\ref{Fig4}(a). In contrast, chains with opposite helicities give MZMs of opposite chirality, which can be coupled, as schematically illustrated in Fig.~\ref{Fig4}(b). 

The topological character of such coupled-chain systems can also be directly computed using a real-space Bott index (equivalent to a winding number for 1D systems with chiral symmetry) \cite{Luobott}:
\begin{equation}
N = \frac{1}{4\pi i} \operatorname{Tr}\!\Bigl[ C \log\!\bigl( M Q M^{\dagger} Q \bigr) \Bigr],
\label{eqn3}
\end{equation}
where \(M = e^{2\pi i x / L_x}\) (with \(L_x\) the length of the system along \(x\)), $P=\sum_n |\psi_n\rangle\langle\psi_n|$ is the projector onto the occupied states, and \(Q = 1 - 2\sum_n |\psi_n\rangle\langle\psi_n|\), and \(|\psi_n\rangle\) are the occupied eigenstates of the BdG Hamiltonian. This index provides a robust tool for exploring the topological phase diagram even in the presence of disorder or magnetic fluctuations.

Beyond 1D Majorana end modes, the chiral symmetry $C$ can protect 2D higher-order TSCs that host Majorana corner modes. This phase can be realized by a dimerized coupling of superconducting helimagnets with alternating spin helicities, as illustrated in Fig.~\ref{Fig4}(c). Because adjacent chains have opposite helicities, the Majorana end modes of the underlying 1D TSCs (oriented along \(x\)) couple through the inter-chain hopping along \(y\). At the two edges \(x = -L_x/2\) and \(x = L_x/2\), these modes form an effective 1D Kitaev chain along the \(y\) direction. When the intracell hopping \(t_1\) is smaller than the intercell hopping \(t_2\) (\(|t_1| < |t_2|\)), this edge Kitaev chain itself becomes topological and harbors MZMs at its ends. Under full open boundary conditions, this results in four Majorana corner modes, as shown in Fig.~\ref{Fig4}(d). This higher-order TSC is characterized by a Bott index \(\tilde{N}\), obtained from Eq.~\eqref{eqn3} by replacing the unitary matrix \(M\) with \(\tilde{M} = e^{2\pi i x y/(L_x L_y)}\), where \(L_y\) is the length of the system along \(y\) \cite{Luobott}.

\begin{figure*}
\centering
\includegraphics[width=5.in]{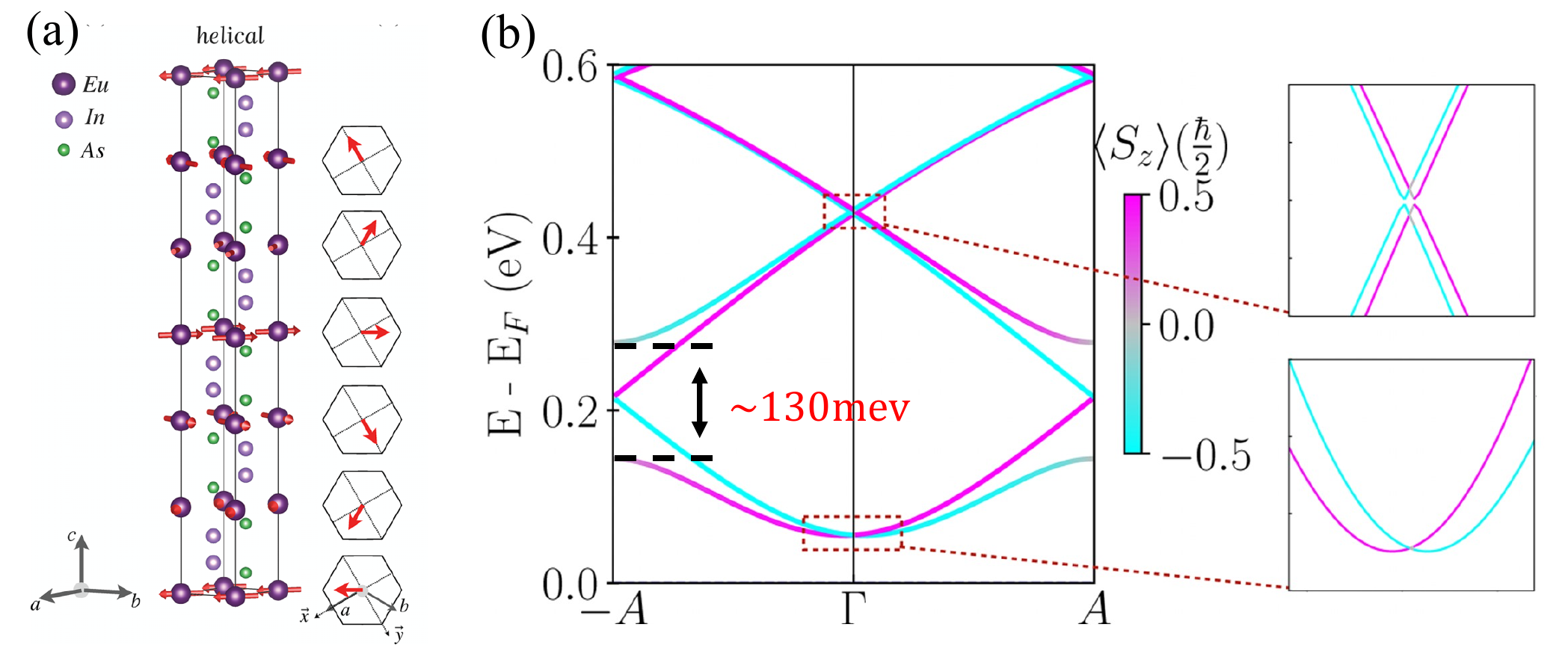}
\caption{ (a) Coplanar magnetic textures of EuIn\(_2\)As\(_2\). (b)  Spin-polarized energy bands without SOC of the helical phase along the -A-A path, where \(\mathrm{A} = (0,0,\pi)\) in reciprocal lattice units. The color encodes the out-of-plane spin expectation value. The small plots on
the right are zoomed-in views near the $\Gamma$ point. The figures are adapted from  Ref.~\onlinecite{k9p4tfhd}} 
\label{Figs5}
\end{figure*}

\subsection{F2. Spin-group-symmetry-protected Majorana Kramers pairs}
Although time-reversal symmetry \(\mathcal{T}\) is explicitly broken in magnetic systems, certain composite anti-unitary symmetries \(\tilde{\mathcal{T}}\) that satisfy \(\tilde{\mathcal{T}}^2 = -1\) can be preserved. Such symmetries place a superconducting system in the DIII symmetry class and can protect Majorana Kramers pairs, allowing a $Z_2$ classification. 

Specifically, we consider two coupled helimagnets with opposite magnetic moments on a superconducting substrate, 
as schematically illustrated in Fig.~\ref{Fig4}(e). This configuration respects the spin group symmetry \(\tilde{\mathcal{T}} = \{\mathcal{T}||M_y\}\), where $M_y$ flips $y$ and $\tilde{\mathcal{T}}^2=-1$. Each helimagnet contributes one Majorana end mode, and the two end modes are related by \(\mathcal{\tilde{T}}\), forming a Kramers pair. This pair remains robust against any perturbation that preserves \(\mathcal{\tilde{T}}\), including a finite out-of-plane magnetization component \(m_z\), as long as the composite symmetry is maintained.
To diagnose the presence of such protected Kramers pairs, we employ a spin-resolved Bott index \(\mathcal{N}\), defined by Eq.~\eqref{eqn3} with the unitary matrix \(M\) replaced by \(\mathcal{M} = e^{2i\pi x \sigma_z / L_x}\) \cite{Luobott}. Although spin is not conserved, this index accurately captures the \(\mathbb{Z}_2\) topology associated with the Kramers pair, as demonstrated numerically in Fig.~\ref{Fig4}(f).

Majorana Kramers pairs can also be realized in a magnetic domain-wall geometry. Figure.~\ref{Fig4}(g) shows a 1D chain with a domain wall at \(x=0\) where the helicity changes sign. This setup respects the spin group symmetry \(\tilde{\mathcal{T}} =\{\mathcal{T}|| M_x\}\) with $M_x$ being the spatial mirror operation along the $x$ direction. Superconducting pairing drives both left and right segments into 1D topological phases, giving rise to two MZMs localized at the domain wall [Fig.~\ref{Fig4}(h)]. These two MZMs are related by \(\tilde{\mathcal{T}}\) symmetry and thus form a protected Kramers pair.

\subsection{G. Band structure of the $p$-wave magnet \texorpdfstring{EuIn\(_2\)As\(_2\)}{EuIn2As2}}

\texorpdfstring{ EuIn\(_2\)As\(_2\)}{EuIn2As2} is a candidate  $p$-wave magnet \cite{2025luospin}, which crystallizes in a hexagonal structure and hosts a helical spin configuration, as shown in Fig.~\ref{Figs5}(a). This compound belongs to the spin space group \(P^{6_{001}^1}6_3 / ^{2_{100}}m^1m{^{6_{001}^1}}c(1,1,3^1_{001})^m1\) \cite{2025luospin} and respects the effective time-reversal symmetry \(\tilde{\mathcal{T}}_1 = \mathcal{T}\sigma_z\) and a spin-group symmetry \(g_2 = \{3_{001}^1||1|\tau\}\), which constrain the spin expectation values to satisfy \(s_{x,y}(\bm{k}) = s_{x,y}(-\bm{k})\) and \(s_z(\bm{k}) = -s_z(-\bm{k})\). The out-of-plane component \(s_z(\bm{k})\) transforms as the \(A_{2u}\) representation of the point group $6/mmm$, with the basis function \(z\), see the
Supplemental Material of Ref.~\onlinecite{2025luospin}. Consequently, the leading term of the NSS in the low-energy expansion can be effectively described by \(k_z\sigma_z\), in agreement with first-principles calculations shown in Fig.~\ref{Figs5}(b).

Moreover, the system respects the spin-group symmetry \(g_3 = \{-1||6_{001}|\tau_{0,0,0.5}\}\), \cite{CheXiaobing2024}. Along the \(\Gamma\)--A line with \(\mathrm{A} = (0,0,\pi)\) in reciprocal lattice units, \(g_2\) acts as an effective time-reversal operation with \(g_3^2 = -e^{ik_z}\). This guarantees Kramers degeneracy at \(k_z = 0\), while the degeneracy at \(k_z = \pi\) can be lifted by the broken time-reversal symmetry, as clearly seen in the first-principles-calculated  bands shown in Fig.~\ref{Figs5}(b). The resulting   magnetic energy gap is approximately \(130\ \mathrm{meV}\). This large intrinsic gap, far exceeding the scale achievable by practical laboratory magnetic fields, exemplifies a key advantage of OPMs as a robust, field-free platform for engineering TSCs.

\end{widetext}

\end{document}